\documentclass[sigconf]{acmart}

\def\BibTeX{{\rm B\kern-.05em{\sc i\kern-.025em b}\kern-.08emT\kern-.1667em\lower.7ex\hbox{E}\kern-.125emX}}
    
\copyrightyear{2019}
\acmYear{2019}
\setcopyright{rightsretained}
\acmConference[CCS '19]{2019 ACM SIGSAC Conference on Computer and Communications Security}{November 11--15, 2019}{London, United Kingdom}
\acmBooktitle{2019 ACM SIGSAC Conference on Computer and Communications Security (CCS '19), November 11--15, 2019, London, United Kingdom}
\acmPrice{}
\acmDOI{10.1145/3319535.3354212}
\acmISBN{978-1-4503-6747-9/19/11}

\usepackage{threeparttable}
\usepackage{booktabs}
\usepackage{subcaption}
\usepackage{tabularx}
\usepackage{balance}
\usepackage{paralist}

\newcommand*\rot{\rotatebox{90}}

\newcommand{\ie}{i.\,e.}
\newcommand{\eg}{e.\,g.}

\begin{document}
\fancyhead{}
%
% The "title" command has an optional parameter, allowing the author to define a "short title" to be used in page headers.
%
% Use https://capitalizemytitle.com/ for title case
% Do not insert line breaks in your title.
\title[(Un)informed Consent: Studying GDPR Consent Notices in the Field]{(Un)informed Consent: Studying GDPR Consent Notices in the Field}

%
% The "author" command and its associated commands are used to define the authors and their affiliations.
% Of note is the shared affiliation of the first two authors, and the "authornote" and "authornotemark" commands
% used to denote shared contribution to the research.

\author{Christine Utz}
\orcid{0000-0003-4346-6911}
\affiliation{%
    \institution{Ruhr-Universit{\"a}t Bochum}
    \city{Bochum}
    \country{Germany}
}
\email{christine.utz@rub.de}

\author{Martin Degeling}
\orcid{0000-0001-7048-781X}
\affiliation{%
    \institution{Ruhr-Universit{\"a}t Bochum}
    \city{Bochum}
    \country{Germany}
}
\email{martin.degeling@rub.de}

\author{Sascha Fahl}
\affiliation{%
    \institution{Ruhr-Universit{\"a}t Bochum}
    \city{Bochum}
    \country{Germany}
}
\email{sascha.fahl@rub.de}

\author{Florian Schaub}
\orcid{0000-0003-1039-7155}
\affiliation{%
    \institution{University of Michigan}
    \city{Ann Arbor}
    \state{Michigan}
}
\email{fschaub@umich.edu}

\author{Thorsten Holz}
\orcid{0000-0002-2783-1264}
\affiliation{%
    \institution{Ruhr-Universit{\"a}t Bochum}
    \city{Bochum}
    \country{Germany}
}
\email{thorsten.holz@rub.de}

%
% By default, the full list of authors will be used in the page headers. Often, this list is too long, and will overlap
% other information printed in the page headers. This command allows the author to define a more concise list
% of authors' names for this purpose.
\renewcommand{\shortauthors}{Utz and Degeling, et al.}

%
% The abstract is a short summary of the work to be presented in the article.
\begin{abstract}
Since the adoption of the General Data Protection Regulation (GDPR) in May 2018 more than 60\,\% of popular websites in Europe display \emph{cookie consent notices} to their visitors. 
This has quickly led to users becoming fatigued with privacy notifications and contributed to the rise of both browser extensions that block these banners and demands for a solution that bundles consent across multiple websites or in the browser. 
In this work, we identify common properties of the graphical user interface of consent notices and conduct three experiments with more than 80,000 unique users on a German website to investigate the influence of notice position, type of choice, and content framing on consent. We find that users are more likely to interact with a notice shown in the lower (left) part of the screen. Given a binary choice, more users are willing to accept tracking compared to mechanisms that require them to allow cookie use for each category or company individually.
We also show that the wide-spread practice of nudging has a large effect on the choices users make. Our experiments show that seemingly small implementation decisions can substantially impact whether and how people interact with consent notices. Our findings demonstrate the importance for regulation to not just require consent, but also provide clear requirements or guidance for how this consent has to be obtained in order to ensure that users can make free and informed choices.
\end{abstract}




% Keywords. The author(s) should pick words that accurately describe the work being
% presented. Separate the keywords with commas.

%
% This command processes the author and affiliation and title information and builds
% the first part of the formatted document.
\maketitle

%-------------------------------------------------------------------------------
\section{Introduction}
%-------------------------------------------------------------------------------
\label{sec:introduction}
In recent years, we have seen worldwide efforts to create or update privacy laws that address the challenges posed by pervasive computing and the ``data economy''. 
Examples include the European Union's General Data Protection Regulation (GDPR) \cite{gdpr_2016}, which went into effect on May 25, 2018, and the California Consumer Privacy Act (CCPA) \cite{ccpa_2018}, which becomes effective on January 1, 2020. These laws uphold informational self-determination by increasing transparency requirements for companies' data collection practices and strengthening individuals' rights regarding their personal data.

The GDPR's impact was twofold. While the number of third-party services on European websites barely changed \cite{sorensen_gdpr_tracking_2019}, websites now ask users for consent prior to setting cookies. In mid-2018, about 62\,\% of popular websites in the EU were found to display a \emph{(cookie) consent notice}, often referred to as ``cookie banner,'' and in some countries an increase of up to 45 percentage points since January 2018 was observed
~\cite{degeling_gdpr_2019}. The design and complexity of such consent notices greatly vary: Some merely state that the website uses cookies without providing any details or options, while others allow visitors to individually (de)select each third-party service used by the website. Paired with the fact that consent notices often cover parts of the website's main content, this high prevalence has led website visitors to become fatigued with consent mechanisms~\cite{burgess_popups}. Consequently, tools have emerged that provide pragmatic workarounds --- one example is the ``I don't care about cookies'' browser extension~\cite{i_dont_care_about_cookies_2019}.
But oftentimes this only leads to data collection taking place without consent since the default on many websites is to employ user tracking \emph{unless} the visitor has opted out~\cite{friedman_privacyux_2019}, and 80\,\% of popular EU websites do not offer any type of opt-out at all \cite{degeling_gdpr_2019}.

Instead of adopting opt-in solutions or enforcing the existing Do-Not-Track standard, the online advertising industry has developed a consent framework~\cite{iab_framework_2019} to reduce the number of consent requests.
Notices using this framework ask website visitors if they consent to data collection for different purposes by up to 400 listed third-party advertisers. Information about their consent decision is then passed down the online advertising supply chain.

Overall, consent notices have become ubiquitous but most provide too few or too many options, leaving people with the impression that their choices are not meaningful and fueling the habit to click any interaction element that causes the notice to go away instead of actively engaging with it and making an informed choice.

Most notice designs only partially use the available design space for consent notices. But we have also seen notices that, \eg, do not force users to accept cookies, ask for consent without hidden pre-selections, or provide visitors with granular yet easy-to-grasp mechanisms to control the website's data processing practices. Hence, we expect that how a consent notice asks for consent has a large impact on how website visitors interact with it, and we are positive that there are design decisions that better motivate people to interact with consent notices in a meaningful way instead of annoying them.

In this paper, we systematically study design properties of existing consent notices and their effects on consent behavior. We systematize consent notices using a sample of 1,000 notices collected from live websites and identify common variables of their user interfaces. 
Our research goal is to explore the design space for consent notices to learn how to encourage website visitors to interact with a notice and make an active, meaningful choice. Over the course of four months, we conduct a between-subjects study with 82,890 real website visitors of a German e-commerce website and investigate their (non-)interaction with variants of consent notices. We collect passive clickstream data to determine how users interact with consent notices and invite them to participate in a voluntary follow-up online survey to obtain qualitative feedback. The study comprises three distinct field experiments to answer the following research questions:

\begin{compactenum}
\item Does the position of a cookie consent notice on a website influence visitors' consent decisions? (Experiment~1, n = 14,135)
\item Do the number of choices and nudging via emphasis / pre-selection influence users' decisions when facing cookie consent notices? (Experiment~2, n = 36,530)
\item Does the presence of a privacy policy link or the use of technical / non-technical language (``this website uses cookies'' vs. ``this website collects your data'') influence users' consent decisions? (Experiment~3, n = 32,225)
\end{compactenum}

In a short follow-up survey answered by more than 100 participants, we ask website visitors to voluntarily report the motivation for their selection, how they perceive the notice they have seen, and how they expect consent notices to function in general.

We find that visitors are most likely to interact with consent notices placed at the bottom (left) position in the browser window while bars at the top of the screen yielded the lowest interaction rates. This is mainly due to the (un)importance of the website content obstructed by the notices and suggests taking into account characteristics of the individual website to identify the notice position most likely to encourage user interaction. 
Interaction rates were higher with notices that provided at most two options compared to those that let users (de)activate data collection for different purposes or third parties individually, even though those notices do not allow visitors to express consent freely. We also show that the more choices are offered in a notice, the more likely visitors were to decline the use of cookies. This underlines the importance of finding the right balance between providing enough detail to make people aware of a website's data collection practices and not overwhelming them with too many options.
At the same time, nudging visitors to accept privacy-invasive defaults leads more visitors to accept cookies, whereas in a privacy-by-default (opt-in) setting, less than 0.1\,\% of visitors allow cookies to be set for all purposes. This suggests that the current data-driven business models of many webservices, who often employ dark patterns to make people consent to data collection, may no longer be sustainable if the GDPR's data protection by default principle is enforced.
Technical language (``This site uses cookies'' instead of ``This site collects your data'') appears to yield higher interaction rates with the consent notice but decreases the chance that users allow cookie use. We find that the presence of a link to the site's privacy policy does not increase user interaction, underlining the importance of making information immediately actionable rather than pointing to further resources.

Survey feedback indicates that users favor category-based choices over a vendor-based approach, and they expressed a desire for a transparent mechanism. A common motivation to give consent is the assumption that the website cannot be accessed otherwise.

Based on the results of our field study, we conclude that opt-out consent banners are unlikely to produce intentional/meaningful consent expression. We therefore recommend that websites offer opt-in notices based on categories of purposes. Above all, we observed that the majority of website visitors does not accept cookies for all purposes, and feedback from our survey suggests that a unified solution that does not interfere with every single website yet provides more control than a simple yes--no decision would best fit users' needs.

%-------------------------------------------------------------------------------
% \section{Background}
\section{Consent Notices}
%-------------------------------------------------------------------------------
\label{sec:background}
We first describe the legal background of consent notices and current challenges for their practical implementation. Then we identify and analyze variables of the user interface of commonly used types of consent notices.

\subsection{Background}
%-----------------------------------

Cookie consent notices emerged in the wake of the European Union's Directive 2009/136/EC \cite{cookie_directive_2009}. 
The directive changed Article 5(3) of the ePrivacy Directive (2002/58/EC) \cite{eprivacy_directive_2002} to require that data is stored on users' devices only after having obtained user consent based on \emph{``clear and comprehensive information [...] about the purposes of the processing.''} An exemption to this consent requirement is storing of information that is \emph{``strictly necessary,''} such as session or authentication cookies. 

On May 25, 2018, the European Union's General Data Protection Regulation (GDPR; Regulation (EU) 2016/679) went into effect. Its Article 6 contains six legal bases for the processing of personal data of European residents, including that \emph{``the data subject has given consent to the processing of his or her personal data for one or more specific purposes''}. Recital 32 of the GDPR and guidelines published by EU data protection authorities \cite{article_29_gdpr_consent} require for valid consent \emph{``a clear affirmative act''} that is a \emph{``freely given, [purpose-] specific, informed and unambiguous indication of [...] agreement to the processing of personal data.''}
Another document clarifies the relationship between the ePrivacy Directive (2002/58/EC) and the GDPR for the use of cookies: Article 5(3) of the directive governs access to non-necessary cookies in the user's browser, whether it contains personal data or not, while the GDPR applies to subsequent processing of personal data retrieved via cookies \cite{edpb_gdpr_eprivacy_2019}.

Degeling et al. found that after the GDPR went into effect 62.1\,\% of 6,579 popular websites in Europe displayed cookie consent notices, compared to 46.1\,\% in January 2018 \cite{degeling_gdpr_2019}.

This high prevalence has sparked efforts to reduce the number of consents required. The most widely used solution, supported by the online advertising industry, is the Transparency \& Consent Framework by IAB Europe \cite{iab_framework_2019}. This framework has been criticized for its bundling of purposes~\cite{ryan_iab_2018} and a lack of transparency regarding the parties the website visitor's personal data could be shared with~\cite{degeling_gdpr_2019, ryan_iab_2018}. An October 2018 decision by the French data protection authority CNIL~\cite{cnil_consent_2018} pointed out a lack of consent verification in the framework, and in April 2019 a formal complaint was filed against the IAB for showing a consent notice on its own website that forces visitors to consent if they want to access the website~\cite{ryan_complaint_2019}, which is not allowed under GDPR.

Another suggestion to decrease the number of consent prompts is to move consent decisions to the browser and let users locally specify their data collection preferences \cite{oneill_dnt_gdpr_2018}. The browser then sends adequate signals to the websites requesting data collection. This would require websites to respect the opt-out signals requested by the browser --- something that has not worked out in the past with the Do-Not-Track standard~\cite{mayer_DNT_2012}.

\subsection{Properties of Consent Notices}
%-----------------------------------

Consent notices currently found on websites vary both in terms of their user interface and their underlying functionality. Regarding the latter, Degeling et al. identified distinct groups within existing implementations of consent notices~\cite{degeling_gdpr_2019}. Some are only capable of displaying a notification that the website uses cookies or collects user data without providing any functionality to make the website comply with the visitor's choice. In contrast, other cookie notices are provided by third-party services that offer complex opt-in choices and block cookies until the user consents explicitly. 

\begin{table*}[h!tb]
	\centering
	\small
	\caption{Variables of the user interface of consent notices and their values across a sample of 1,000 drawn from 5,087 consent notices collected from the most popular websites in the European Union in August 2018}
	\label{tab:cb_properties}
	\begin{threeparttable}
		\begin{tabular}{@{}lrlrlrlrlr}
			\toprule
			\textbf{Position} & & \textbf{Choices (visible)}          &  & \textbf{Choices (hidden)} & & \textbf{Blocking} & & \textbf{Nudging} & \\
			top          & 27.0\,\% & no option    & 27.8\,\% & no option    & 26.3\,\% & yes &  7.0\,\% & yes          & 57.4\,\%\\
			bottom       & 57.9\,\% & confirmation & 68.0\,\% & confirmation & 59.9\,\% &  no & 93.0\,\% & no           & 14.8\,\%\\
			top right    &  0.2\,\% & binary       &  3.2\,\% & binary       &  4.0\,\% &     &          & n/a\tnote{a} & 27.8\,\%\\
			bottom right &  3.0\,\% & categories   &  1.0\,\% & slider       &  0.2\,\% &\\
			top left     &    0\,\% & vendors      &    0\,\% & categories   &  8.1\,\% &\\
			bottom left  &  3.7\,\% &              &          & vendors      &  1.1\,\% &\\ 
			center       &  7.8\,\% &              &          & other        &  0.4\,\% &\\
			other        &  0.4\,\% &&&&&& \\
			 \midrule
			\multicolumn{2}{l}{\textbf{Link to privacy policy}} & \textbf{Text: Collection} & & \textbf{Text: Processor} & & \textbf{Text: Purposes} & &&\\
            yes   & 92.3\,\% & ``cookies'' & 94.8\,\% & unspecified & 75.5\,\% & generic  & 45.5\,\% &&\\
            no    &  6.6\,\% & ``data''    &  1.4\,\% & first party &  0.7\,\% & specific & 38.6\,\% &&\\
            other &  1.1\,\% &   both      &  1.6\,\% & third party &  2.6\,\& & none     & 16.9\,\% &&\\
                  &          &   none      &  0.9\,\% & both        & 21.1\,\% &          &          &&\\
                  &          &   other     &  1.3\,\% & other       &  0.1\,\% &          &          &&\\
			\bottomrule
		\end{tabular}

	\centering
	\begin{tablenotes}
		\item[a] Nudging is not available for ``no option'' notices.
	\end{tablenotes}

	\end{threeparttable}
\end{table*}

Our study focuses on the \emph{user interface} of consent notices, a topic which has not been systematically studied before. In order to identify common properties of consent notices currently used on websites, we analyze a random sample of 1,000 notices drawn from a set of 5,087 we collected in a previous study \cite{degeling_gdpr_2019}. To obtain that set, the following steps were taken: First we created a list of websites containing the 500 most popular websites for each member state of the European Union as identified by the ranking service Alexa~\cite{alexa}. This yielded a list of more than 6,000 unique domains. Using a Selenium-based automated browser setup, we visited all of them in an automated way in August 2018 from an IP address within the EU and took screenshots of each website's home page. We then manually inspected these screenshots if they contained a consent notice. 
In our previous study, we identified six distinct types of choices consent notices offer to website visitors, as described below. In this work, we extend our prior analysis to other variables of the user interface of consent notices. For this, we took the 5,087 consent notices collected previously, drew a random sample of 1,000 notices, and manually inspected how they differed in their user interface. We identified the following eight variables, whose possible values, along with their frequency in our random sample, are listed in Table~\ref{tab:cb_properties}:

\textbf{Size.} 
The size of the consent notice as displayed in the browser. We found the value of this variable to vary widely depending on the implementation of the notice, from small boxes that only cover a fraction of the viewport to notices taking up the whole screen. Responsive web design may result in the same notice using up different shares of the viewport, depending on the screen size and orientation of the device used to view the website. Typically notices take up a larger percentage of the viewport on smartphones than on desktop computers and tablets. The size of a consent notice may also be fixed by design, \ie, to cover the whole viewport of any device.

\textbf{Position.}
We observed the consent notices in our dataset to be displayed in seven distinct positions: in one of the four corners of the viewport (\textit{dialog} style; 6.9\,\%), at the top (27.0\,\%) or bottom (57.9\,\%) like a website header or footer (\textit{bar} style), and vertically and horizontally centered in the middle of the viewport (7.8\,\%)).
On smartphones in portrait mode, the limited space reduces the number of options to the top, bottom, and middle of the screen.

\textbf{Blocking.} 
Some consent notices (7.0\,\%) prevent visitors from interacting with the underlying website before a decision is made~\cite{schaub_design_2015}. The site's content may also be blurred out or dimmed \cite{friedman_privacyux_2019}. All consent notices shown in the center position were blocking. We also observed some blocking consent notices at the top or bottom position.

\textbf{Choices.} Consent notices offer website visitors different choice options.  We identified the following mechanisms for user interaction~\cite{degeling_gdpr_2019}:
\begin{compactitem}
	\item \textbf{No option} notices simply inform the user that the website uses cookies without any option for interaction. The user continuing to use the website is interpreted as agreement to the notice.
	\item \textbf{Confirmation-only} banners feature a button with an affirmative text such as ``OK'' or ``I agree'', clicking on which is interpreted as an expression of user consent.
	\item \textbf{Binary} notices provide two buttons to either accept or decline the use of all cookies on the website.
	\item 
	\textbf{Category}-based notices group the website's cookies into a varying number of categories. Visitors can allow or disallow cookies for each category individually, typically by (un)checking a checkbox or toggling a switch. For transparency reasons, the category of ``strictly necessary'' cookies (whose use does not require consent according to Article 5(3) of Directive 2002/28/EC) is often also listed but the switch to deactivate it is greyed out. Some notices use a \textbf{slider}: Instead of (de)selecting categories individually the user can move a slider to select one of the predefined levels, which implies consent to all of the previously listed categories.
	\item \textbf{Vendor}-based notices offer even more fine-grained control by allowing visitors to accept or decline cookies for each third-party service used by the website. Such notices are part of IAB Europe's Transparency and Consent Framework \cite{iab_framework_2019}, which refers to its advertising partners as ``vendors.''
\end{compactitem}

\textbf{Text.} The text displayed by consent notices also varies widely. It should inform the website visitor of the fact that the website uses cookies or similar tracking technology and may list additional information such as the purpose of the data collection. Depending on the choices offered, the notice may provide instructions for consenting to (or denying) the use of cookies. Table~\ref{tab:cb_properties} provides an overview of common text contents of consent notices for the following typical pieces of information: 
\begin{compactitem}
    \item \textbf{Collection.} What the visitor consents to, which can be the use of cookies (94.8\,\%), the collection of their personal data (1.4\,\%), both (1.6\,\%), neither (0.9\,\%), or something else (such as the website's privacy policy; 1.3\,\%).
    \item \textbf{Processor.} Who collects this information, which can be specifically limited to the first party (0.7\,\%), third-party services (2.6\,\%), both (21.1\,\%), or refer to an unspecified party (usually denoted by the pronoun ``we'' or the domain/website name; 75.5\,\%).
    \item \textbf{Purposes.} These may be specific (\eg, ``audience measurement'' or ``ad delivery''; 38.6\,\%), generic (\eg, ``to improve user experience''; 45.5\,\%), or not specified at all (16.9\,\%).
\end{compactitem}

\textbf{Nudging \& Dark Patterns.} Consent notices often (57.4\,\%) use interface design to steer website visitors towards accepting privacy-unfriendly options. Typical techniques include color highlighting of the button to accept privacy-unfriendly defaults, hiding advanced settings behind hard to see links, and pre-selecting checkboxes that activate data collection \cite{forbrukerradet_deceived_2018}. We observed all of these techniques in our sample.

\textbf{Formatting.} We found that, unless predetermined by the consent library used, the choice of fonts and colors typically matched that of the underlying website. The formatting of consent notices may also be influenced by the website's business requirements \cite{friedman_privacyux_2019}, \eg, sites relying on monetization via online behavioral advertising (OBA) are unlikely to steer their visitors towards an opt-out mechanism by making this option highly visible.

\textbf{Link to additional information.} Consent notices may include a link to the website's privacy policy, a designated cookie policy, or a website providing additional information about cookies -- 92.3\,\% of the notices in our sample contain such a link to additional information. In Table~\ref{tab:cb_properties}, we marked as ``other'' consent notices where the full privacy policy was already included in the notice itself (1.1\,\%).

Table \ref{tab:cb_properties} shows that the majority of consent notices are placed at the bottom of the screen (58\,\%), not blocking the interaction with the website (93\,\%). They offer no options besides a confirmation button that does not do anything (86\,\%), and most try to nudge users towards consenting (57\,\%). While nearly all notices (92\,\%) contain a link to a privacy policy, only a third (39\,\%) mention the specific purpose of the data collection or who can access the data (21\,\%).

%-------------------------------------------------------------------------------
\section{Method}
%-------------------------------------------------------------------------------
\label{sec:method}
Given the legal requirements for explicit, informed consent, the vast majority of cookie consent notices we analyzed are likely not compliant with European privacy law. To further investigate the effects of different combinations of these properties on consent behavior, we conducted a field study with consent notices on a German e-commerce website.

We investigated the effect of the following parameters on users' interactions with consent notices:
\begin{compactenum}
    \item The \emph{position} of the notice, as notices displayed in some parts of the screen are more likely to be ignored.
    \item The number of \emph{choices} offered by the notice, which is influenced by legal requirements and the need to give users actual control over the website without overwhelming them with too many options.
    \item \emph{Nudging} visitors towards giving consent through highlighting and preselection, since this may cause people to consent who would not have made the same decision otherwise.
    \item The presence of a \emph{privacy policy link} and whether the notice refers to ``cookie use'' (technical language) or ``data collection'' (non-technical language). These differences in wording may influence people's expectations of the website's data processing practices and thus their consent decision.
\end{compactenum}

We did not evaluate the effects of the following parameters: blocking (because the owner of our partner website asked us not to block access to the site), formatting (because of the multitude of options -- we chose the same color scheme as in the notice previously used on the website), and size (which is difficult to vary consistently across devices). 

From the end of November 2018 to mid-March 2019, we conducted three between-subjects experiments to determine if, and how, different parameters of consent notices influence interaction rates. In each experiment, we tested variants for one or two of the parameters described in Table~\ref{tab:cb_properties}: position in Experiment~1, choices and nudging in Experiment~2, and wording and the presence of a privacy policy link in Experiment~3. The respective other parameters were kept constant in an experiment.

\subsection{Study Setup}
We partnered with a German-language e-commerce website based on WordPress. 
The website has 15,000--20,000 unique visitors per month, most of which are single-page visitors that reach the site from a search engine looking for product information and reviews. 
The third-party services used by the website are Google Fonts and the CSS framework Ionic for design, Google Analytics embedded via Google Tag Manager for audience measurement, Facebook social media buttons, embedded YouTube videos, and targeted advertisements delivered by Google Ads. All of these services store cookies in the visitor's browser.

We modified a WordPress plugin, Ginger -- EU Cookie Law \cite{manafactory_ginger_2018}, to test different notice variants. Ginger was selected because it can block cookies before opt-in, log users' consent, and because it was released under a GPLv2 license. By the time of publication of this paper, the original version of the plugin had been discontinued. 
We added support for checkbox-based and ``no option'' notices. We did not implement ``slider'' notices because we considered them a less compliant variant of the ``categories'' type. 

The plugin was further modified to function as follows in our study: When a user first visited our partner website, they were shown one consent notice. Which notice of the $n$ test conditions in the current experiment was displayed was determined in round-robin fashion. The ID of the displayed notice was stored in a cookie in the participant's browser to ensure visitors who did not click the notice would continue to see the same notice across subpages and recurring visits. Each participant was assigned a unique identifier: $pid$$=$$\text{SHA-256}(ip\_address || user\_agent)$. The participant's IP address was discarded after computation of $pid$. The participant ID was stored in another cookie, together with the participant's consent as required by Article 7 GDPR\footnote{The legal bases for storing the cookie that remembers the banner ID are Article 6(1)(e) GDPR (public interest in conducting this study) and Article 6(1)(c) GDPR (compliance with a legal obligation) for storing the consent cookie.}.

If the visitor clicked any interaction element that would usually cause a consent notice to disappear, \ie, the `X' discard button, ``Accept,'' ``Decline,'' or ``Submit,''\footnote{In all experiments, all texts in the consent notice and survey were in German to match the website's language. Survey responses were also in German. The authors translated all texts and responses into English for this paper. Both the original and the translated consent notices and the survey are available in our GitHub repository at \url{https://github.com/RUB-SysSec/uninformed-consent}.} the notice did not disappear instantly. Instead, the notice content was replaced with an invitation to take an online survey about their experiences with this and other consent notices (see Appendix~\ref{sec:survey-responses}). The invitation disclosed that this was a university study and that participants could win one of 15 25-euro shopping vouchers. Users could either click ``Discard'' to close the notice, or select ``Participate'' to open the survey in a new browser tab. The survey was created in a LimeSurvey instance running on a web server hosted by the authors.

If the website visitor did not interact with the consent notice, the content of the notice was automatically replaced with the survey invitation 30 seconds after the page had fully loaded. This is because we also wanted to explore users' reasons for \emph{not} interacting with consent notices. Web analytics data for our partner website showed that 95\,\% of all users who had interacted with the website's previous consent notice had done so within 30 seconds of accessing the site. Thus we assumed that website visitors who did not interact with the consent notice within 30 seconds would not have clicked it at a later point in time. 

\begin{figure*}[tb]
	\centering
	\includegraphics[width=1.0\textwidth]{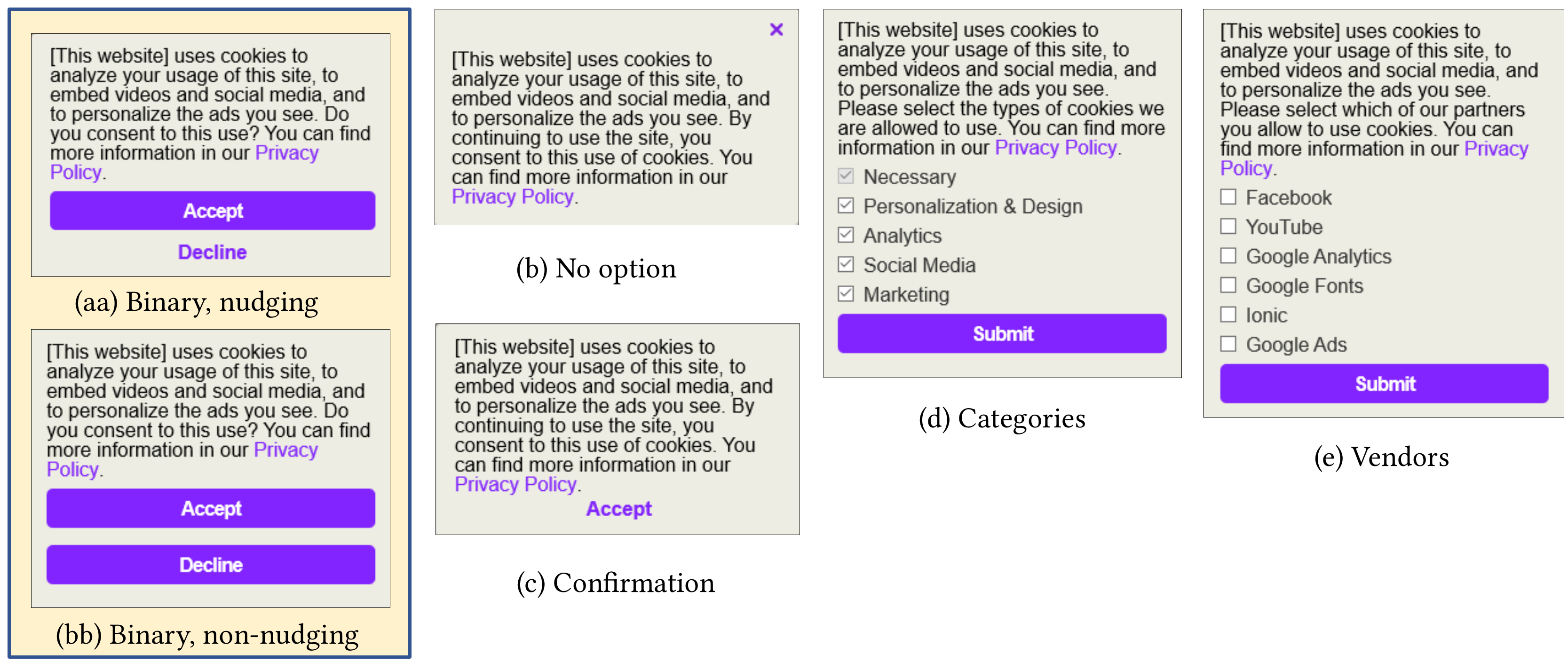}
	\caption{Cookie consent notices with different choice mechanisms and nudging used in our experiments: (a) a binary notice in two variants, one nudging visitors to click ``Accept'' (aa) and one presenting both choices equally (bb); (b) a no-option notice (nudging not applicable); (c) a confirmation-only notice (shown without nudging); (d) a category-based notice with pre-selected checkboxes (nudging); and (e) a vendor-based notice with unchecked checkboxes (non-nudging).}
	\label{fig:cookie_banners}
\end{figure*}

\begin{figure}[t]
\includegraphics[width=0.4\textwidth]{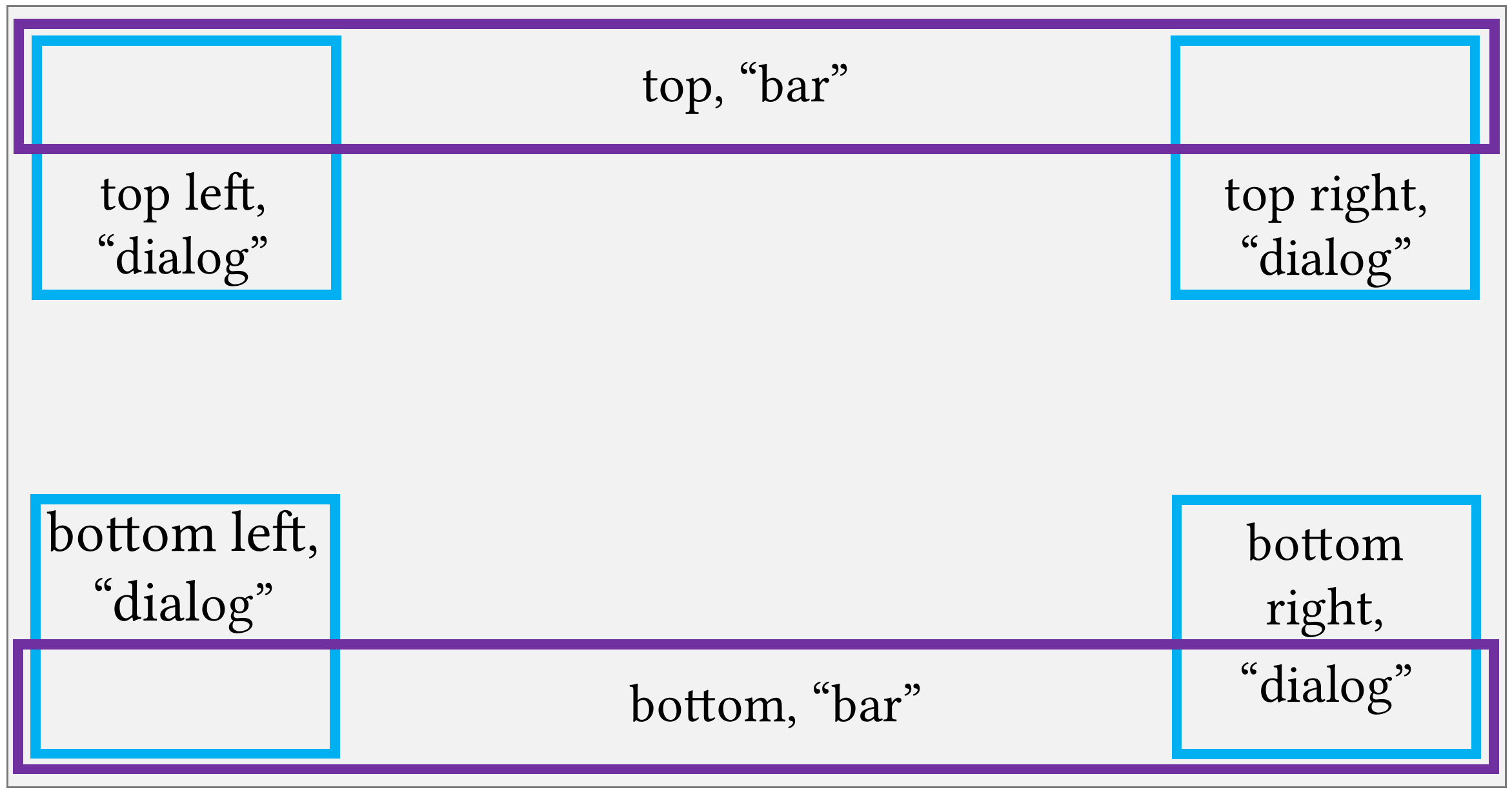}
\caption{Positions tested in Experiment~1.}
\label{fig:positions}
\end{figure}

We modified the Ginger plugin's logger add-on to create log entries whenever a participant clicked an interaction element on the notice. 
Log events were also triggered upon page load, when links to the privacy policy or survey were clicked, when the consent notice content was auto-replaced with the survey invitation, and when the participant dismissed this invitation.
Each log entry consisted of a timestamp, the participant's ID ($ pid $), the ID of the consent notice they had seen, the event they had triggered, their screen resolution, operating system, browser, and whether an ad blocker had been detected.\footnote{We used BlockAdBlock 3.2.1 (\url{https://github.com/sitexw/BlockAdBlock}) to detect ad blocking functionality in the visitor's browser.}

\subsection{Experiment 1: Position}
%-----------------------------------

Experiment 1 ran from November 30 to December 18, 2018, \ie, for 19~days. We had observed consent notices being shown at various screen positions and wanted to determine the effect of placement on interaction with the cookie consent notice to inform our subsequent experiments. The research question for Experiment~1 was: \emph{Does the cookie consent notice's position on a website influence a visitor's consent decision?} In order to encourage user interaction, we displayed a ``binary'' notice without nudging (see Figure~\ref{fig:cookie_banners}(bb)), the simplest type offering an actual choice. We tested the notice in six different positions (see Figure~\ref{fig:positions}).
We could not test the center position as our partner asked us to not block access to their website.

\subsection{Experiment 2: Number of Choices, Neutral Presentation vs. Nudging}
%-----------------------------------

From December 19, 2018 to January 28, 2019, we conducted Experiment~ 2, which focused on the effects of given choices and pre-selections on consent. In our analysis of consent notices, we had identified various complexity levels of choices offered and methods to emphasize certain options. Prior work has shown that the design and architecture of choices heavily influences people's decisions~\cite{weinmann_nudging_2016, thaler_nudge_2009}. While this effect has also been shown successful in improving user privacy \cite{acquisti_nudging_2009, acquisti_nudging_2017}, in practice it is most often used to make users share more information~\cite{forbrukerradet_deceived_2018}.
Website owners often have an interest in getting visitors to agree to the use of cookies and hence highlight certain choices in the consent notice to nudge visitors towards accepting. We observed this for 57.4\,\% of the notices in our sample. Our research question therefore was: \emph{Does the number of choices and nudging through emphasis or pre-selection in consent notices influence user's consent decisions?}

For nudging, we used pre-checked checkboxes and buttons highlighted in contrasting colors, techniques often used to nudge users towards accepting default settings \cite{forbrukerradet_deceived_2018}. While we observed that most category- and vendor-type notices in practice display such fine-grained controls only after the visitor clicked ``Settings,'' we chose to immediately display all available options to ensure that our conditions only varied in the number and framing of choices. 
In Experiment~2, we displayed the following consent notices at the position determined in Experiment~1 to yield the highest interaction rates:

\begin{compactitem}
	\item \textbf{No option} (Figure~\ref{fig:cookie_banners} (b)): 
	In line with many notices we observed, we added an `X' in the top-right corner to dismiss the banner. There is no nudging variant because the notice does not offer any choice.
	\item \textbf{Confirmation--Non-nudging} (Figure~\ref{fig:cookie_banners} (c)): This notice has an ``Accept'' button which is not highlighted.
	\item \textbf{Confirmation--Nudging}: Same as the Confirmation--Non-nudging notice, but the ``Accept'' button is highlighted (like the ``Accept'' button in Figure~\ref{fig:cookie_banners} (a) (aa)). 
	\item \textbf{Binary--Non-nudging} (Figure~\ref{fig:cookie_banners} (a) (bb)): The ``Accept'' and ``Decline'' buttons are formatted the same way, neither is emphasized.
    \item \textbf{Binary--Nudging} (Figure~\ref{fig:cookie_banners} (a) (aa)): Same as Binary--Non-nudging but only the ``Accept'' button is highlighted in a contrasting color. 
	\item \textbf{Categories--Non-nudging}: Same as notice (d) in Figure~\ref{fig:cookie_banners}, but with unchecked checkboxes. The ``Necessary'' category cannot be unchecked, as is common practice. 
	\item \textbf{Categories--Nudging} (Figure~\ref{fig:cookie_banners}(d)): Same as Categories--Non-nudging but with pre-checked checkboxes for all categories.
	\item \textbf{Vendors--Non-nudging} (Figure~\ref{fig:cookie_banners}(e)): Similar to the categories variant, but the checkboxes correspond to the third-party services used by our partner website.
	\item \textbf{Vendors--Nudging}: Same as Vendors--Non-nudging but with pre-selected checkboxes.
\end{compactitem}

For the category-based notices, we had to map the third-party services used by the website to different categories. We manually inspected the 434 category-based notices in our initial set of 5,087 consent notices for common category wording. For example, we found advertising cookies to be categorized as ``marketing'' or ``advertising''; web analytics was also referred to as ``performance cookies,'' ``statistics,'' or ``audience measurement.'' This yielded the following category--third party mappings:

\begin{compactitem}
    \item \textit{Necessary:} Cookies to remember the displayed notice and the website visitor's consent decision.
    \item \textit{Personalization \& Design:} Ionic, Google Fonts
    \item \textit{Analytics:} Google Analytics
    \item \textit{Social Media:} Facebook, YouTube
    \item \textit{Marketing:} Google Ads
\end{compactitem}

For all category- and vendor-based notices in Experiments~2 and 3, the available options were displayed in random order, except for the ``Necessary'' category, which was always displayed first as in the majority of category-based notices we had observed.

In Experiments~2 and 3, we increased the font size of the banner message, resulting in larger notices. We did this to fix an implementation bug of the Ginger plugin that had caused the text to be displayed in a very small font on some smartphones in portrait mode.

\subsection{Experiment 3: (Non-)Technical Language and Privacy Policy Link}

Experiment~3 was conducted from January 29 to March 15, 2019. In this experiment, we tested the influence of the presence of a link to the website's privacy policy. Previous research suggests that (American) Internet users have consistent misconceptions about privacy policies, indicated by the fact that a majority believes the existence of a privacy policy means that a website cannot share personal data with third parties~\cite{turow_persistent_2018}. At the same time, Martin~\cite{martin_privacy_2016} showed that the existence of a reference to a privacy policy in the context of data sharing explanations increases mistrust in a website. 
There are further known misconceptions about what cookies actually are and what they are used for~\cite{mcdonald_cookies_2010, ha_cookie_perception_2006}.
To learn more about the influence of these factors in the context of consent notices, our research question was: \emph{Does the presence of a privacy policy link or mention of cookies influence users' consent decisions?}

The base notice for this experiment was the Category--Non-nudging notice from Experiment~2 because of GDPR's data protection by default requirement and the ability to provide consent for specific purposes with checkboxes. We chose a category-based notice over a vendor-based one due to the results of Experiment~2 (see Section~\ref{sec:study2}).
The notice text for this experiment was: 
``This website [uses cookies | collects your data] to analyze your usage of this site, to embed videos and social media, and to personalize the ads you see. Please select for which purposes we are allowed to use your data. [You can find more information in our privacy policy].'' We tested the following conditions:

\begin{compactitem}
    \item \textbf{Technical--PP Link:} The original Categories--Non-nudging notice from Experiment~2. It uses both technical language (``uses cookies'') and a sentence with a link to the website's privacy policy.
    \item \textbf{Technical--No PP Link:} Same as Technical--PP Link, but the privacy policy sentence was replaced with whitespace to keep the size of the notices consistent.
    \item \textbf{Non-Technical--PP Link:} Same as Technical--PP Link, but using non-technical language (``your data'' instead of ``cookies''). 
    \item \textbf{Non-Technical--No PP Link:} Same as Non-Technical--PP Link, but with the privacy policy sentence replaced with whitespace.
\end{compactitem}

For participants who saw a notice with non-technical language, we replaced other occurrences of the term ``cookie'' in our setup: In the study invitation, ``cookie notice'' was replaced with ``privacy notice,'' and we adjusted the wording of some survey questions and response options as described in Appendix~\ref{sec:survey-responses}.

%-------------------------------------------------------------------------------
\subsection{Research Ethics}
%-------------------------------------------------------------------------------
\label{sec:ethics}
Our study was conducted on a website with real users, which raises ethical concerns as we did not ask for consent prior to measuring their interactions with consent notices. We did so to ensure ecological validity and be able to capture non-biased results as we expected the majority of visitors to not pay attention to a study consent notice asking them to opt in, which was supported by our findings.

While our institution does not require IRB review for minimal risk studies, we ensured that we did not deceive or harm website visitors and their privacy.
All displayed consent notices functioned as described and respected the visitor's choice.  To test the effect of no-option consent notices, we had to offer fewer choices than we believe is required by the GDPR. We added a paragraph describing our study to the website's privacy policy.
The data we collected was pseudonymized. Logs were stored on the website's server and access was limited to two researchers conducting the analysis and the website's owner. After the study, the data was removed from the server and copied to the researchers' data center.

All visitors were informed about the study after 30 seconds when we showed a notice asking them for participation in the survey. Survey participants were asked for explicit consent and to confirm they were over 18 and wanted to participate. Email addresses of participants who opted to participate in the prize draw were stored separately from the dataset, without the participant ID.

%-------------------------------------------------------------------------------
\subsection{Data Analysis}
%-------------------------------------------------------------------------------
\label{sec:data-analysis}
\subsubsection{Event logs}
When we started the data analysis, we noticed inconsistencies in some entries. 
The event logs created by our plugin indicated that some website visitors had seen multiple notice versions. This could have happened because users had deactivated cookies completely, visited the website in multiple sessions using private browsing mode, or opened the website in multiple tabs simultaneously. For another set of users, we detected multiple screen resolutions, mostly because the screen orientation had changed. Rotating the screen could lead to the notice covering different parts of the website, so we removed these participants to preserve consistency. In total, we removed 2,1\,\% of participants across all experiments.

\subsubsection{Survey}
We considered a survey response complete if the participant had at least answered Q1--Q6 but did not provide a free-text answer to Q7 and Q8.
Due to a low survey response rate we received few responses for some conditions. We therefore refrained from a quantitative analysis of survey responses.
In Section~\ref{sec:evaluation}, we evaluate responses to the open-ended questions (parts of Q1; Q6--Q8). We coded these responses using emergent thematic coding. Two of the authors independently devised a set of codes for each question and coded the responses. The results were discussed and yielded a final codebook, which was used to re-code all responses. Any remaining disagreements were reconciled by the two coders. We report the codes and their distribution in Appendix~\ref{sec:survey-responses}, along with the answers to all closed-ended questions.

%-------------------------------------------------------------------------------
\section{Results}
%-------------------------------------------------------------------------------
\label{sec:evaluation}
\subsection{Dataset and Website Visitors}

Our cleaned dataset contained event logs of 82,890 unique website visitors: 14,135 in Experiment~1, 36,530 in Experiment~2, and 32,225 in Experiment~3. 21.72\,\% of all visitors accessed the website on a desktop or laptop computer and 78.28\,\% with a mobile device (of which 5.1\,\% were tablets)\footnote{We count as ``desktop computer'' actual desktop machines as well as laptops. ``Mobile'' devices include smartphones and tablets; the latter were used by 5.1\,\% of visitors.}.
Overall, 6.95\,\% of participants used an ad blocker. The rate was much higher on desktop (29.1\,\%) than on mobile devices (0.8\,\%). These numbers are consistent with a 2017 report for Germany~\cite{pagefairadblock}, the highest rate of ad block users in Western Europe (20\,\% on average), and North America (18\,\% on average). 
For 16.45\,\% of visitors, we could not detect whether they used an ad blocker. These visitors did not stay long enough on the website to complete ad blocker detection.
On average, users spent a short time on the website. Pre-study Google Analytics data provided by the partner website showed that 84.81\,\% of visitors spend less than 10 seconds on the site, 5.21\,\% 11 to 60 seconds, and 5.83\,\% up to 3 minutes. Our dataset includes all users for whom the event logs indicated a fully loaded site, regardless of how long they stayed on the page, resulting in a high number of ``no action'' visitors. As described in Section~\ref{sec:study2}, the median time until an interaction with any version of the notice was 4 to 8 seconds. About 11,800 users stayed on the page for 10 seconds or more.

The link to our survey was clicked 804 times (168 in Experiment 1, 445 in Experiment 2, and 191 in Experiment 3). We received a total of 110 responses (16 in Experiment~1, 60 in Experiment~2, and 34 in Experiment~3), which means that 0.37\,\% of the 29,712 visitors who interacted with the notice or stayed on the site for longer than 30~seconds participated in the survey. 
To get an impression of visitors' expectations about the website's data collection practices, we asked \emph{Q2: What do you think -- what data does [the website] collect about you when you access the website?} This question was answered by all participants. Across all three studies, the data most commonly expected to be collected were links clicked on the site (78\,\%), IP address (65\,\%), posts read on the site (61\,\%), and the device used (59\,\%). Less often mentioned were other sites visited (29\,\%) and the visitor's place of residence (25\,\%). 13\,\% thought the website collected their name, even though the site never asks for it. Only 5\,\% thought the site did not collect any data about them. 
These answers indicate that the survey participants had a good understanding of what data websites can collect even without user accounts.

\subsection{Experiment 1: Banner Position}
\begin{figure}[t]
	\centering
    \includegraphics[width=0.4\textwidth]{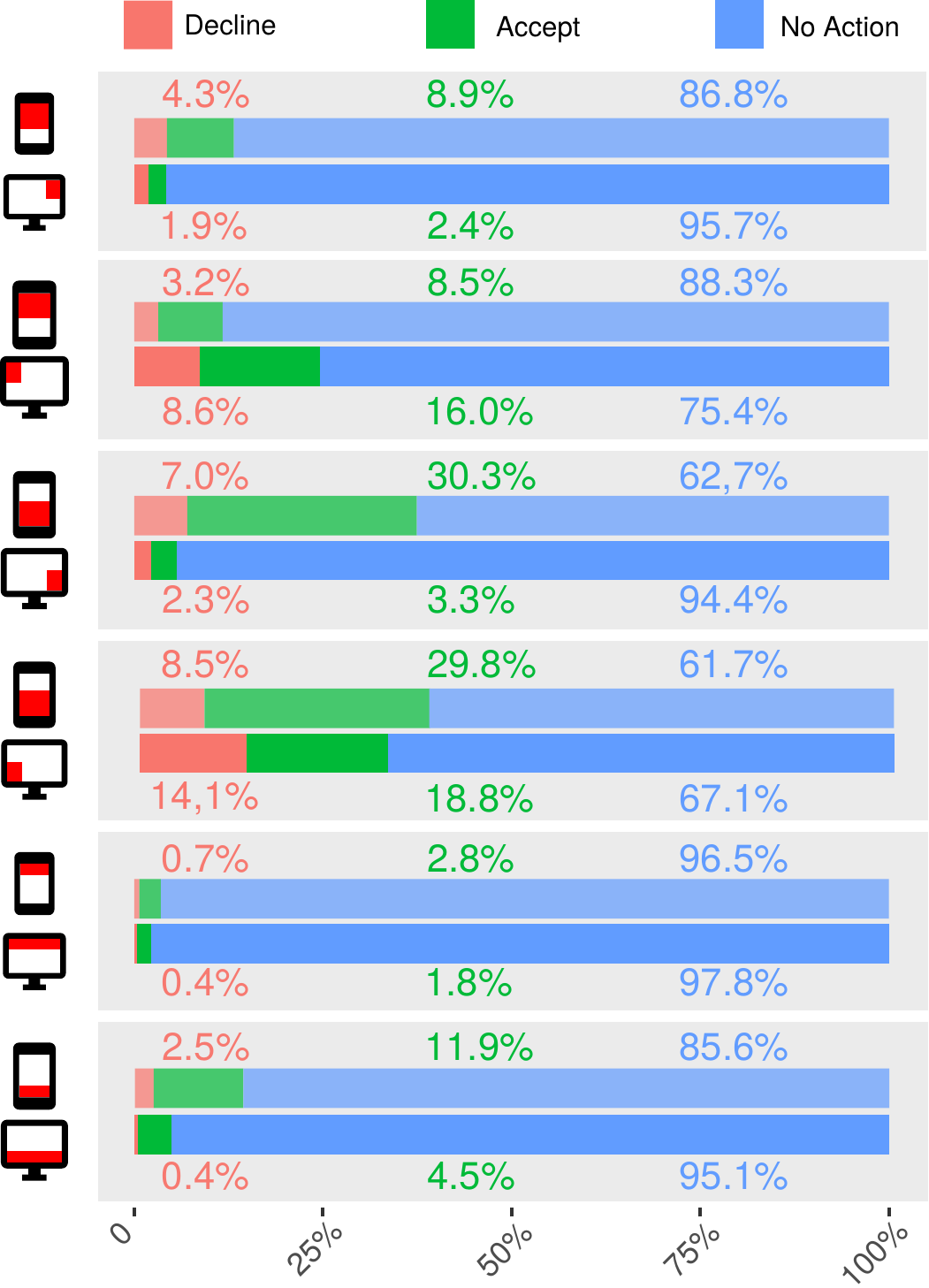}
     \caption{Interaction rates in Experiment~1 (notice position), arranged pairwise for mobile and desktop users.}
	\label{fig:study1-events}
\end{figure}

\subsubsection{Interaction rates}
Figure~\ref{fig:study1-events} shows how visitors interacted with the consent notices displayed at different positions. % on the website. 
Overall the notices shown at the bottom-left position received the most interactions, 37.1\,\% of visitors interacted with them regardless of device type or choice made. 
The notice positions most commonly observed in practice, small bars at the top or bottom, resulted in low interaction (2.9\,\% and 9.6\,\%, respectively).

While we were mainly interested in position in Experiment~1, we also analyzed the influence of other variables, such as ad blocker use, screen resolution, browser, operating system, and device type (desktop/mobile). We estimated the effect size of different properties by calculating Cram\'{e}r's~V (CV) and over all visitors the banner position showed the largest effect size (CV=.31). Unless noted otherwise, $\chi^2$-tests for effects in this experiment are statistically significant ($p$$<$$.001$).

Ad blocker use also had a small impact on whether someone interacted with the notice. While on average 15.8\,\% of visitors without an ad blocker interacted with any notice, only 12.6\,\% of ad blocker users did so, but the effect size was rather small (CV=.11). The impact of screen resolution was much higher on desktop (CV=0.33) than on mobile (CV=0.16): Only 5.5\,\% of visitors with screen resolutions of 1,920 by 1,080 pixels or higher interacted with the notice, while the average was 25.6\,\% for smaller screens. Although the decline/accept ratio varied between conditions, we could not identify a single factor to explain the differences. Across all conditions the number of users who accepted cookies was higher than the number of those that declined.

\subsubsection{Discussion}

A possible explanation for higher interaction rates with notices displayed at the bottom is that these notices are more likely to cover the main content of the website, while notices shown at the top mostly hide design elements like the website header or logo. If one uses their thumb to navigate websites on a smartphone, it is also easier to tap elements on the bottom part of the screen than those at the top. An explanation for higher interaction rates with notices displayed on the left of the viewport might be the left-to-right directionality of Latin script: Line breaks cause the information density of a text to be skewed to the left, so consent notices positioned on the left are more likely to obstruct visitors' reading and trigger an interaction with the notice.

We looked for qualitative feedback in the survey responses. In Experiment~1, we received 16 responses, with eight participants having interacted with the notice and another eight that did not. 
All six participants who answered they had clicked the notice ``because it prevented them from reading the website content'' had seen a notice shown at the bottom or left side. 

Both on desktop and mobile, the notice positioned in the bottom-left corner received the most attention. 
Thus, we decided to display the notices in Experiments~2 and 3 in the bottom-left corner.

\subsection{Experiment 2: Choices \& Nudging}
\label{sec:study2}
In Experiment~2 there were 36,395 participants in total.
Each of the nine conditions was shown to 4,044 website visitors on average.

\subsubsection{Interaction rates}

\begin{figure}[tb]
	\centering
    \includegraphics[width=0.45\textwidth]{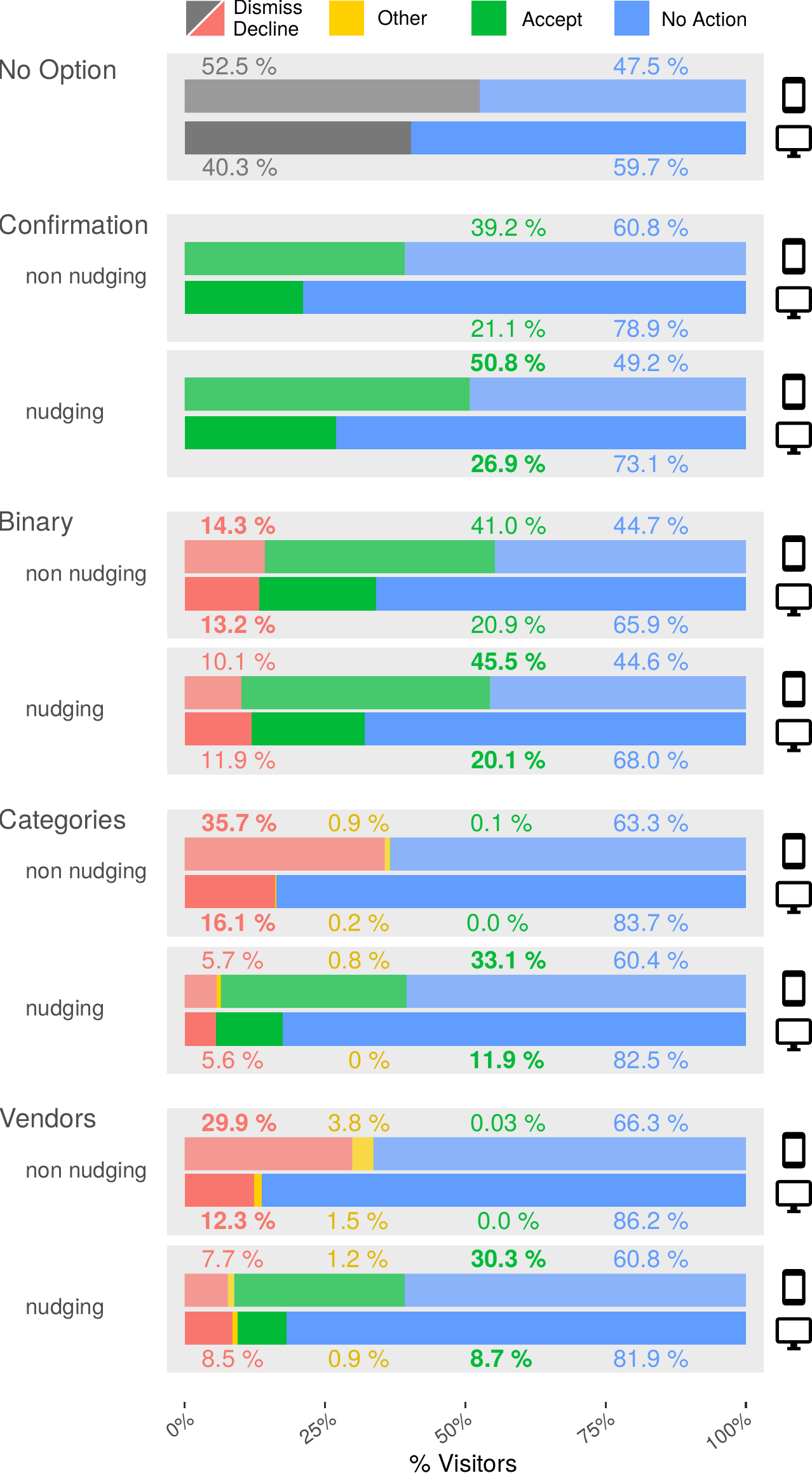}
    \caption{Visitors' consent choices in Experiment~2. ``Accept''/``Decline'' indicate that (all) options were accepted or declined. ``Other'' includes those who accepted/declined only some options. Bold figures indicate default options.}
    \label{fig:events-r2}
\end{figure}

Figure~\ref{fig:events-r2} provides an overview of the recorded visitor interactions. Compared to Experiment~1, the overall percentage of visitors who interacted with the notice increased (13,8\,\%--55,3\,\%), especially on mobile devices, likely because we had increased the font size, resulting in larger notices. The highest interaction rate (55\,\%) was measured for binary notices on mobile devices. 

The experiment revealed a strong impact of nudges and pre-selections. Overall the effect size between nudging (as a binary factor) and choice was CV=.50. For example, even for confirmation-only notices, more users clicked ``Accept'' in the nudge condition, in which it was highlighted (50.8\,\%  mobile, 26.9\,\% desktop), than in the non-nudging condition, in which ``Accept'' was displayed as a text link (39.2\,\% m, 21.1\,\% d). The effect was most pronounced for category- and vendor-based notices, in which all checkboxes were pre-selected in the nudging conditions, but not in the privacy-by-default conditions.
The pre-selected versions led around 30\,\% of mobile users and 10\,\% of desktop users to accept all third parties. In contrast, only a small fraction (< 0.1\,\%) allowed all third parties when given the opt-in choice and 1 to 4\,\% allowed one or more third parties (``other'' in Figure~\ref{fig:events-r2}), indicating that some users still engaged with the offered choices. No desktop visitors allowed all categories. 
Interestingly, the number of non-interacting users was highest on average for the vendor-based conditions, although they took up the largest amount of screen space due to six options being offered. We discuss qualitative survey feedback on the category- and vendor-based notices in Section~\ref{sec:survey-choice}.

\subsubsection{Choices}

Results were mixed in terms of the consent choices users made when given options (in all but the no-option and confirmation conditions). Surprisingly, more participants accepted cookies in both binary conditions, where they had the option to decline cookies, than in the non-nudging confirmation condition, where they could only accept cookies or not interact with the notice.

\begin{figure*}[t]
	\centering

	\includegraphics[width=1\textwidth]{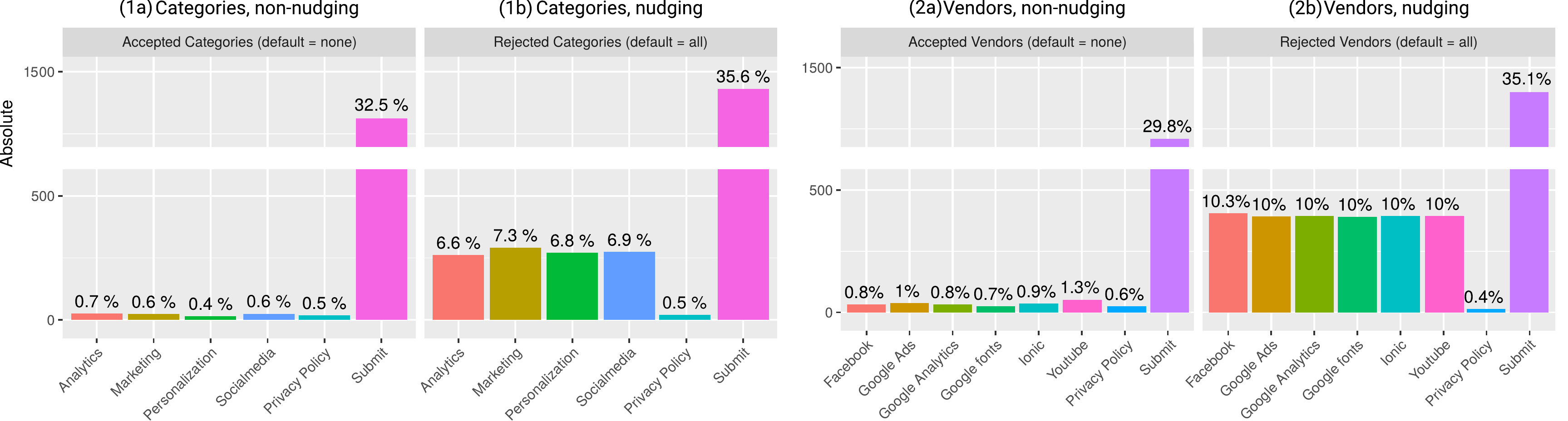}

	\caption{Decisions to allow or decline specific categories (1) or vendors (2) in the the specific conditions of Experiment~2. Subgraphs (a) show how many visitors checked specific boxes, subgraphs (b) how many unchecked pre-selected boxes.}
	\label{fig:decision-events}
\end{figure*}

Figure \ref{fig:decision-events} lists the specific choices participants made on category- and vendor-based notices. Few visitors chose specific categories or vendors if they were not pre-selected (non-nudging conditions). Interestingly, more visitors selected specific vendors than  categories. Vendors YouTube and Ionic were selected most, even though survey responses (Q6) indicated that Ionic was lesser known than other listed vendors. 
We observe a similar pattern for the de-selection of specific categories and vendors: More visitors unchecked one or more vendors (10.0\,\%) than categories (6.9\,\%).

6\,\% of visitors who saw a category- or vendor-based notice clicked at least one of the checkboxes more than once. 48 visitors (0,08\,\%) toggled an even number of times, reversing previous decisions. Interestingly, 47 of those users saw a ``nudging'' notice so that they actively reactivated one of the categories. 

We also recorded how long it took visitors to submit their choice.
The median time to submit for no-option, confirmation and binary-choice notices was 4--5 seconds; 7--8 seconds for category- or vendor-based notices.\footnote{We report the median as the data showed a high standard deviation since we had no way to check when the interaction with a notice started, and sometimes the choice was submitted minutes after the page had been loaded.} For details see Appendix~\ref{sec:timing}.

\subsubsection{External validation}
\label{sec:validity}

To verify the generalizability of our results, which are only based on visitors to our partner website, we compared our data to internal data from Cookiebot, a company offering cookie consent notices (similar to our category-based conditions) as a service to websites. 
Their dataset from February 2019 contains 3 million user logs for 2,000 different websites. The Cookiebot notices also show purpose categories, so we compare their data with our data for the category-type notices. In their case, some of the checkbox selections cannot be changed by users, as website owners can argue that the use of certain cookie categories is based on different legal grounds (\eg, ``legitmate interest'', Art. 6 (1) (f) GDPR). Therefore (de)selecting all consent-based cookie categories in Cookiebot notices sometimes requires fewer clicks to be made, and we were not able to compare decisions we labeled as ``other''.
As shown in Table~\ref{tab:consentcompany}, Cookiebot has a slightly higher acceptance rate (5.6\,\% compared to 0.16\,\% in our dataset) and a lower decline rate when all boxes are pre-selected (1.2\,\% compared to 16.5\,\% in our dataset). This means that our findings are generally comparable, but specific results may differ based on website and category, which is what we would expect given that privacy preferences are highly contextual \cite{acquisti_behavior_2015}. A related 2017 study (n = 300) found that about 3\,\% of users are willing to accept marketing cookies~\cite{ryan_consent_2017}, which is between marketing acceptance in our non-nudging (0.6\,\%) and nudging  (7.3\,\%) conditions.

\begin{table}[t]
    \centering
    \caption{Comparison of interactions with category notices}
    \label{tab:consentcompany}
        \begin{tabular}{llll}
            \toprule
            Dataset  &  Decision & None pre-selected  & all pre-selected\\
            \midrule 
            Cookiebot & & (n = 1,135,090) & (n = 1,988,681)  \\
            & Accept & 5.59\,\% & 98.84\,\% \\
            & Decline & 94.41\,\% & 1.16\,\% \\
		
            Our Data & & (n = 1,239) & (n = 1,380) \\
            & Accept & 0.16\,\% & 83.55\,\% \\
            & Decline & 99.84\,\% & 16.45\,\% \\
            \bottomrule
    \end{tabular}
\end{table}

\subsubsection{Discussion}
Experiment~2's results show that nudges and pre-selection had a high impact on users' consent decisions. 
It also underlines that the GDPR's data protection by default requirement, if properly enforced, could ensure that consent notices collect explicit consent. We further find that most visitors make binary decisions even when more choices are offered by agreeing to all or no options. Only very few visitors selected specific categories or vendors, while even in the non-nudging binary condition a considerable number accepted the use of cookies. An explanation for this behavior might be that those who are somewhat OK with cookie use are not willing to expend effort on enabling it. Another explanation, suggested by previous work~\cite{martin_privacy_2016}, is that showing the actual practices decreases the trust in a website and therefore leads to more users making an informed decision to decline cookies.

\subsection{Experiment~3: Language \& Privacy Policy Link}

\begin{figure}[t]
	\centering
    \includegraphics[width=0.45\textwidth]{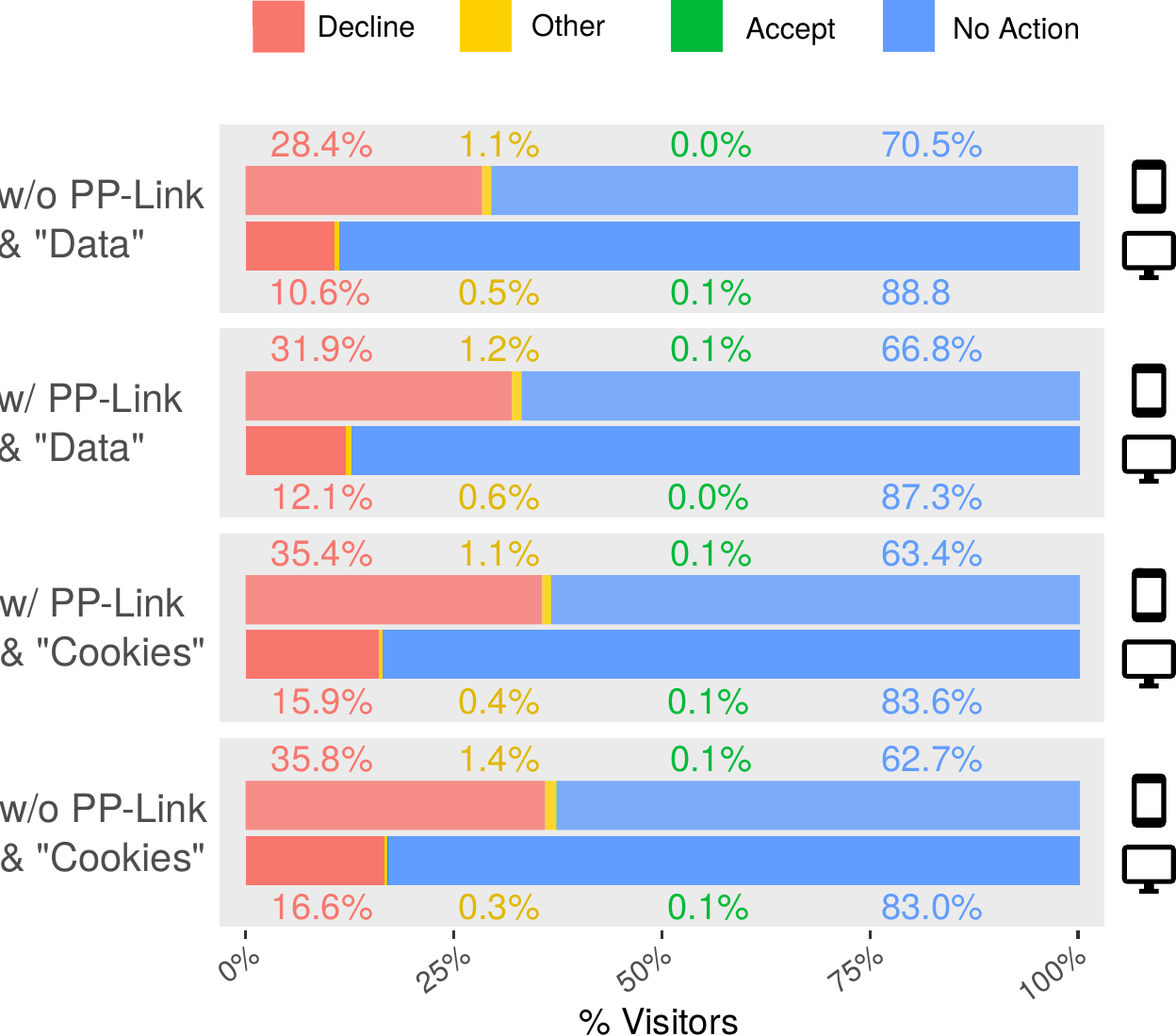}

    \caption{Visitors' interactions with different consent mechanisms in Experiment~3. Notices contained technical language (``cookies'') and a link to the privacy policy (or not).}
    \label{fig:events-r3}
\end{figure}

In Experiment~3, we tested four conditions with combinations of (a) the notice including a link to the privacy policy (or not) and (b) the text either referring to ``cookies'' or ``your data'' more generally. All conditions were variants of the category-based, non-nudging notice from Experiment~2. Figure~\ref{fig:events-r3} summarizes the results. All conditions were shown to 6,032 visitors on average. Again,  interaction rates were higher for mobile visitors. As in Experiment~2, very few visitors accepted all categories (0--0.1\,\%), but some visitors (0.3--1.4\,\%) explicitly allowed one or more. 
More people make a choice when technical language is used, \ie, ``cookie'' is mentioned in the notice. While this difference is significant ($\chi^2$-test, ($p$$<$$.01$), the effect size is low (CV=.08), as are the differences between conditions. Presence of the privacy policy link had no significant effect ($p$$<$$.08$).

\subsubsection{Discussion}
Experiment~3 showed that mentioning of cookies has a minor influence on users' consent behavior. However, differences between conditions are small. This is not surprising given that most users either submit the default choice or do not interact with the notice at all. We could not confirm previous studies~\cite{martin_privacy_2016} that showed a negative effect on trust in a website when a privacy policy was mentioned, but we found that more visitors decline the use of cookies if a privacy policy is linked. Our findings indicate that position and choice have a more pronounced effect on consent behavior than notice language or pointers to more privacy information.

\subsection{Survey Results}

\subsubsection{Reasons for (Non-)Interaction with Notices}
\label{sec:survey-motivation}

In the survey (see Appendix~\ref{sec:survey-responses}), we asked participants why they did or did not click on the consent notice. Participants could select multiple reasons.

44 of 61 survey participants who had clicked the notice reported they had done so because they were annoyed by it. 16 thought the website would not work otherwise, and 13 stated they had clicked the notice out of habit. 11 participants interacted with the notice to protect their privacy, 6 for security reasons, and 5 to see fewer ads. 

49 participants had not interacted with the consent notice, 20 of which reported they had not seen the notice. Nine thought clicking the notice would not have any effect, six did not care what cookies the website used or what data it collected, and three thought it did not offer enough choices. Two reported to not know what cookies were or what data the question was referring to. 
13 participants selected ``other'' and provided a free-text response. Recurring themes in these responses include that the notices were ``annoying [...], so I just ignore them out of frustration'' (Participant 2-94)\footnote{The first digit in our participant identifiers denotes the experiment and the second the response ID assigned by LimeSurvey.} and that participants thought no cookies would be set if they did not interact with the notice. One participant mentioned that they ``[found] all of the partners suspicious'' (2-255). One had opened the website in a background browser tab, so they had only seen the invitation to take the survey, and two participants reported that the notice had been auto-replaced before they could click it.

\subsubsection{Perception of Complex Consent Notices}
\label{sec:survey-choice}

We asked survey participants who saw a category- or vendor-based notice to elaborate on their choice selection (Q6), in order to learn how they perceived purpose-based consent mechanisms as required by the GDPR. We received 38 responses across Experiments~2 and 3. Appendix~\ref{sec:survey-responses} lists the codes and their distribution for this and the following open-response questions.

A recurring theme in the responses was transparency, as mentioned by 5 participants who had seen a category-based notice: ``[I liked] that I could directly select the options without going to the settings. It would be great if this was the default'' (3-171), ``What I like [here] is that only [the ...] necessary option is selected and all of the others are deactivated'' (3-88). One participant with a vendor-based notice stated: ``Having options makes me feel secure'' (2-619).

However, participants had diverging opinions regarding the notices' clarity. 
Some found the categories ``self-explanatory'' (3-118).
Others pointed out that ``Necessary [from a technical perspective] does not say much. Cookies aren't necessary to view a website'' (3-215) and that ``something could be hidden'' (2-557) behind the Necessary category. 
6 (of 7) participants who saw a vendor-based notice in Experiment~2 reported it had ``too much text, too many options. I'm interested in the website's content, not in the consent notice'' (2-116), and one suggested ``it would be perfect to have a button to (de)activate all cookies'' (2-199). 
Seven participants based their choices on privacy considerations: ``I don't tick anything. I only need advice [from] the website'' (3-108), ``I don't want personalized web pages, ads, [... and] pointers to social media'' (3-165).

These responses indicate that more complex notices are not necessarily problematic, as long as options are not pre-selected. While some express concerns, do not trust the categorizations, or find the choices too complex, others appreciate the privacy-by-default approach.

\subsubsection{Understanding of Consent Notice Behavior}
\label{sec:questionnaire-understanding}

The survey further investigated participants' general understanding of how consent notices work and what it meant to accept or decline cookies. This section was identical in all three studies. 
The participant was shown the binary notice depicted in Figure~\ref{fig:cookie_banners} (a) (bb). Then we asked the following two free-text questions:
\emph{Q7: What do you think happens when you click ``Decline''?}
\emph{Q8: What do you think happens when you click ``Accept''?}

\subsubsection{Declining Cookies}

For Q7 (Decline), we received 94 responses across the three studies. 
We identified ten themes. 
The most prominent expectation was that declining cookies would prevent access to the website (28 responses): ``I don't get access to the desired information'' (1-282), ``The site closes itself and you are redirected to the search engine'' (2-685).
17 other participants expected parts of the website not to work: 
``I won't be able to use some functionality because [...] cookies fund the website'' (2-255).
Only 4 participants explicitly mentioned that they would be able to access the site, stating, for example, ``Normally I can continue to navigate the site. It has only happened twice that [a] site has kicked me out. But online shopping [is] difficult if you don't agree'' (2-94).
3 participants expected no collection or processing of personal data to take place when cookies are declined but still had doubts: ``I hope that no data is collected'' (1-177, 1-121, 3-216).
12 expected the site to behave as if ``Accept'' were clicked: ``I guess my data is still collected'' (1-170), ``Nothing, of course. Me not accepting cookies does not mean that the site uses less or no cookies or does not collect any data about me'' (2-630).
Other recurring themes in the responses include the expectation to see less ads, a focus on the technical aspects (``no cookies are evaluated'' [3-217]), and if the notice would dis- or reappear. See Appendix~\ref{sec:survey-responses} for details.

\subsubsection{Accepting Cookies}

For Q8 (Accept), which was also answered by 94 participants (not all the same respondents as for Q7), we also identified 10 themes. 
29 participants expected their personal data would be collected and/or processed: ``my behavior on the website is stored and analyzed'' (2-216), ``my data is shared with who knows what third parties [...] Facebook, Google, marketing / market research / ad analytics [...]'' (2-557). 
19 responses focused on technical aspects: 
``a cookie is set which recognizes me when I revisit the website'' (1-250). 
21 participants stated the website would only work if they allowed cookies: ``I can read the article'' (2-53), ``I can continue to use the website'' (2-405). 
Other themes included effects on the consent notice only (``the banner disappears'' [2-675]), personal data being collected for advertising, user profiling, and other purposes, \eg, ``sale to third parties'' (3-171), ``influencing Internet algorithms'' (1-269), and ``any purpose'' (1-207, 3-64). 7 participants believed it made no difference what was clicked but did not specify what that ``default'' behavior of the website would be.

These answers indicate that our participants had some understanding of how cookies are used, \eg, to recognize recurring visitors and for ad tracking and targeting. 
Concerningly, almost a quarter of participants thought they had to accept cookies before they could access a website -- negative experiences on some sites may be influencing general expectations and behavior across websites. A transparent and GDPR-compliant consent notice should inform users which website functionality may not work as intended if cookies are declined.

%-------------------------------------------------------------------------------
\section{Related Work}
%-------------------------------------------------------------------------------
\label{sec:relatedwork}
Multiple measurement studies of varying scope have provided insights about the prevalence of consent notices \cite{article_29_wp_cookie_sweep_2015, degeling_gdpr_2019, vaneijk_location_2019}. Even though many consent notice libraries can be configured to only display a notice to EU visitors~\cite{degeling_gdpr_2019}, van Eijk et~al.~\cite{vaneijk_location_2019} found that a website's top-level domain was the primary factor in whether a consent notice was displayed rather than a visitor's location.

Sanchez-Rola et al.~\cite{sanchez-rola_optout_2019} evaluated the functionality of consent notices and opt-out mechanisms under GDPR. They manually visited 2,000 popular websites, tried to opt out of data collection whenever possible, and studied the effects on the website's cookies. They found that 92\,\% of websites set at least one high-entropy cookie before showing any kind of notice. Only 4\,\% of notices provided an opt-out choice, and 2.5\,\% of websites removed some cookies upon opt-out. Degeling et al.~\cite{degeling_gdpr_2019} further found that many third-party consent libraries either lack the functionality to block or delete cookies, or require significant modification of a website to properly react to visitors' consent choices.

In Section~\ref{sec:background}, we presented a detailed analysis of variants in consent notices' user interfaces. Previous work had only classified consent notices by the provided information~\cite{kulyk_cookie_perceptions_2018}, the choices offered~\cite{degeling_gdpr_2019, sanchez-rola_optout_2019}, and if the notice blocks access to the website~\cite{sanchez-rola_optout_2019}. Van Eijk et~al.~\cite{vaneijk_location_2019} report some statistics on the height and width of consent notices, their location offset, and notices' word and link/button counts.

Kulyk et al.~\cite{kulyk_cookie_perceptions_2018} investigated users' perceptions of and reactions to differently worded cookie consent notices. 
They identified five categories of disclaimers based on the amount of information provided about the purposes of cookie use and the parties involved. In a qualitative user study, they found that the text of a cookie notice does not significantly influence users' decisions to continue using a website; their decision was rather based on the website's perceived trustworthiness and relevance.  
The participants perceived cookie consent notices as a nuisance or threat to their privacy, and reported lacking information about the implications of cookies and possible countermeasures.

Users' perceptions of consent notices' choice architectures have only been partially studied before.
Boerman et al.~\cite{boerman_behavior_2018}, using Dutch panel data, explored how users protect their online privacy. Given the opportunity to decline cookies, many participants self-reported that they decline cookies ``often'' (16\,\%) or ``very often'' (17\,\%). Facing the decision to either accept cookies or leave the website, 12\,\% and 13\,\% reported to refrain from using the site ``often'' and ``very often,'' respectively.

Previous work has shown that cookies are poorly understood by Web users. 
Ha et al. \cite{ha_cookie_perception_2006} studied the usability of two cookie management tools in focus groups, identifying misconceptions about cookies and risks associated with them. Kulyk et al.~\cite{kulyk_cookie_interface_2018} developed and tested a privacy-friendly cookie settings interface for the Chrome browser and found that users appreciate tools that help them better understand the standard browser cookie settings, such as an assistant that transforms users' privacy preferences into cookie settings or additional explanations about the purpose and security/privacy implications of different types of cookies.

Consent notices are not the only way for Web users to opt out of targeted advertising. 
Previous work has evaluated the usability of different opt-out tools \cite{leon_optout_2012, habib_optout_2019, garlach_adchoices_2018} and found that users find it difficult to locate, configure, and understand these mechanisms. 

Schaub et al. describe the design space for privacy notices and controls, including consent notices and permission prompts on mobile devices  \cite{schaub_design_2015}.

Warning research and ad placement studies provide insights into the effects of user interface design choices on user attention and behavior; examples include color \cite{silic_colour_2016} and position \cite{cantoni_positioning_2013}. Studies investigating different notice designs were conducted, for example, for SSL~\cite{felt_ssl_2015}, browser security~\cite{reeder_browser_warnings_2018}, and phishing warnings~\cite{egelman_phishing_2008}.

Mathur et al.~\cite{mathur_patterns_2019} classified common dark patterns in web services. In their classification scheme the observed actions are described as ``sneaking'' (attempting to misrepresent user actions, or delay information that, if made available to users, they would likely object to), ``misdirection'' (using visuals, language, or emotion to steer users toward or away from making a particular choice), and ``forced action'' (forcing the user to do something additional in order to complete their task).

%-------------------------------------------------------------------------------
\section{Discussion}
%-------------------------------------------------------------------------------
\label{sec:discussion}
We conducted three experiments evaluating the effects of cookie consent notices' position, choices, and content on people's consent behavior. In the following we describe recommendations based on our findings and discuss limitations of our approach. 

\subsection{Recommendations}
Our experiments investigated different notice positions, details of the choices offered, and the wording of cookie consent notices. Future guidelines for consent notices should consider the following recommendations:

\paragraph{Position}
Experiment~1 showed that the position of a notice has a substantial impact on whether a website visitor engages with the notice. A dialog box in the lower left corner (on desktop) or the lower part of the screen (on mobile) significantly increases the chance that a user makes a consent decision. While we had expected higher interaction rates on mobile devices for this position since it is easy to reach with the thumb, we were surprised by the impact on desktop users, given the general wisdom that content in the top left receives the most attention in cultures with left-to-right writing.
This result could be related to our partner website, like many websites, displaying a header which shifted content to lower parts of the screen.
This experiment shows that the second most common notice position observed in practice, the top position (see Table~\ref{tab:cb_properties}), results in notices being ignored by users.

\paragraph{Choices}
Our results from Experiment~2 showed that nudging (highlighting ``Accept'' buttons or pre-selecting checkboxes) substantially affects people's acceptance of cookies, providing clear evidence for the interference of such dark patterns with people's consent decisions. 
Given a binary choice, more visitors accepted cookies than declined them, which could be evidence for the adverse effects of consent bundling on consent decisions, which is not allowed under the GDPR. Surprisingly, rejection rates in the vendor- and cookie-based conditions were close to those in the binary condition, although visitors had to make five to six additional clicks to reach the same goal. This suggests that people who want to decline cookies are willing to expend extra effort.

Moreover, the survey answers show that participants think that no data is collected \emph{unless} they make a decision, showing that privacy by default is the expected functionality, although this is not the current practice.

\paragraph{Text}
While we did not see an effect in Experiment~3 from including a privacy policy link in the notice, we found that mentioning ``cookies'' made more users reject the data collection. 
The negative effect of mentioning cookies can very well be related to the fact that Internet users have in general a negative feeling about them~\cite{kulyk_cookie_perceptions_2018, ha_cookie_perception_2006}.

It is clear that the current ecosystem of mechanisms to prompt for user consent --- with a plethora of combinations regarding the provided information, the granularity of user options, and how and if their choice is enforced --- provides no real improvement for user privacy compared to pre-GDPR times. At the same time many things are still in flux, with regulators publishing differing guidelines on how to obtain consent, the online advertising industry developing and updating proposals for consent frameworks, and legal and technical scholars evaluating them. While some claim~\cite{ryan_iab_2018} that many underlying principles of the online advertising industry are not compatible with the GDPR at all, the regulation so far has only partially affected how companies process personal data~\cite{urban_perspectives_2019}.
We hope that our results can inform future discussions, not only with recommendations for the design of consent notices. Given that at the moment very few users are willing to give consent to any form of processing of their personal data, we think that the business model of online behavioral advertising, which targets ads based on large amounts of personal data, should be challenged and alternative models like privacy-friendly contextual advertising or other ways of monetization for web services need to be developed.

%-------------------------------------------------------------------------------
\subsection{Limitations}
%-------------------------------------------------------------------------------
\label{sec:limitations}
Our study has some potential limitations. First, our sample is biased as we conducted all experiments on a German-language e-commerce website whose visitors may not be representative of the general public. 
However, our partnership with this website gave us control over the notice implementation and access to a high number of unique visitors. We validated some of our results with data from Cookiebot which showed similar results (see Section \ref{sec:validity}). Overall it seems our sample is more inclined towards rejecting cookies. We have to assume that in general a higher percentage of users may allow cookies.
Our field study did not allow us to collect more detailed information about visitors, such as their specific device, the size of the notice on the screen, or how long they stayed on the website, which could potentially have an effect on consent behavior. 

Furthermore, many visitors did not interact with the notice at all and spent only a short period of time on the site. While this could be related to the notice, it is not unusual that most visitors leave a site after a few seconds. Liu et al.~\cite{liu_dwell_2010} showed that website dwell time has a negative aging effect. Users first skim a site to decide whether they will stay on it. Since we were not able to measure the exact time visitors stayed on the site, we included all users for whom the logged data indicated a fully loaded page, which results in a high number of ``no action'' visitors. From a legal perspective the time spent on the site does not affect the need to request consent.
Our partner website also does not have user accounts. Past research has shown that visitors tend to underestimate the amount of personal data collected by websites on which they do not create an account and enter personal data \cite{rao_mismatched_2016}. This may cause them to underestimate the privacy implications of allowing cookie use, but we did not see evidence for this in the survey responses.

Responses to our voluntary survey are likely biased due to participants' self-selection. Responses to the question about possible data collection suggest that participants had a good understanding of the technical background or an interest in privacy. Of the survey participants, 61 had previously interacted with our consent notices and 49 had not, showing that the results are only partially biased towards those who care about notices. We considered this bias when interpreting results.

%-------------------------------------------------------------------------------
\section{Conclusion}
%-------------------------------------------------------------------------------
\label{sec:conclusion}
We conducted the first large-scale field study on the effect of cookie consent notices on people's consent behavior. Cookie notices have seen widespread adoption since the EU's General Data Protection Regulation went into effect in May 2018.
Our findings show that a substantial amount of users are willing to engage with consent notices, especially those who want to opt out or do not want to opt in to cookie use. At the same time, position, offered choices, nudging, and wording substantially affect people's consent behavior. Unfortunately, many current cookie notice implementations do not make use of the available design space, offering no meaningful choice to consumers. Our results further indicate that the GDPR's principles of data protection by default and purposed-based consent would require websites to use consent notices that would actually lead to less than 0.1\,\% of users actively consenting to the use of third-party cookies.

% The acknowledgments section is defined using the "acks" environment (and NOT an unnumbered section). This ensures
% the proper identification of the section in the article metadata, and the consistent spelling of the heading.
%
\begin{acks}
The authors would like to thank the owner of their partner website for allowing them to display different sets of consent notices on this site. Additional thanks to Yana Koval for her help with the implementation of the WordPress plugin and the classification of existing consent notices.
This research was partially funded by the MKW-NRW Research Training Groups SecHuman and NERD.NRW, the German Research Foundation (DFG) within the framework of
the Excellence Strategy of the Federal Government and the States (EXC~2092 CaSa -- 39078197), and the National Science Foundation under grant agreement CNS-1330596.
\end{acks}

%
% The next two lines define the bibliography style to be used, and the bibliography file.

\bibliographystyle{ACM-Reference-Format}
\balance 
\bibliography{bibliography}


\begin{thebibliography}{50}

%%% the Web address.

\ifx \showCODEN    \undefined \def \showCODEN     #1{\unskip}     \fi
\ifx \showDOI      \undefined \def \showDOI       #1{#1}\fi
\ifx \showISBNx    \undefined \def \showISBNx     #1{\unskip}     \fi
\ifx \showISBNxiii \undefined \def \showISBNxiii  #1{\unskip}     \fi
\ifx \showISSN     \undefined \def \showISSN      #1{\unskip}     \fi
\ifx \showLCCN     \undefined \def \showLCCN      #1{\unskip}     \fi
\ifx \shownote     \undefined \def \shownote      #1{#1}          \fi
\ifx \showarticletitle \undefined \def \showarticletitle #1{#1}   \fi
\ifx \showURL      \undefined \def \showURL       {\relax}        \fi
% invisible to TeX
\providecommand\bibfield[2]{#2}
\providecommand\bibinfo[2]{#2}
\providecommand\natexlab[1]{#1}
\providecommand\showeprint[2][]{arXiv:#2}

\bibitem[\protect\citeauthoryear{Acquisti}{Acquisti}{2009}]%
        {acquisti_nudging_2009}
\bibfield{author}{\bibinfo{person}{Alessandro Acquisti}.}
  \bibinfo{year}{2009}\natexlab{}.
\newblock \showarticletitle{{Nudging Privacy: The Behavioral Economics of
  Personal Information}}.
\newblock \bibinfo{journal}{\emph{IEEE Security \& Privacy}}
  \bibinfo{volume}{7}, \bibinfo{number}{6} (\bibinfo{date}{Dec.}
  \bibinfo{year}{2009}), \bibinfo{pages}{82--85}.
\newblock
\showISSN{1558-4046}
\urldef\tempurl%
\url{https://doi.org/10.1109/MSP.2009.163}
\showDOI{\tempurl}


\bibitem[\protect\citeauthoryear{Acquisti, Adjerid, Balebako, Brandimarte,
  Cranor, Komanduri, Leon, Sadeh, Schaub, Sleeper, Wang, and Wilson}{Acquisti
  et~al\mbox{.}}{2017}]%
        {acquisti_nudging_2017}
\bibfield{author}{\bibinfo{person}{Alessandro Acquisti}, \bibinfo{person}{Idris
  Adjerid}, \bibinfo{person}{Rebecca~Hunt Balebako}, \bibinfo{person}{Laura
  Brandimarte}, \bibinfo{person}{Lorrie~Faith Cranor}, \bibinfo{person}{Saranga
  Komanduri}, \bibinfo{person}{Pedro Leon}, \bibinfo{person}{Norman Sadeh},
  \bibinfo{person}{Florian Schaub}, \bibinfo{person}{Manya Sleeper},
  \bibinfo{person}{Yang Wang}, {and} \bibinfo{person}{Shomir Wilson}.}
  \bibinfo{year}{2017}\natexlab{}.
\newblock \showarticletitle{{Nudges for Privacy and Security: Understanding and
  Assisting Users' Choices Online}}.
\newblock \bibinfo{journal}{\emph{Comput. Surveys}} \bibinfo{volume}{50},
  \bibinfo{number}{3} (\bibinfo{date}{Aug.} \bibinfo{year}{2017}).
\newblock
\urldef\tempurl%
\url{https://doi.org/10.2139/ssrn.2859227}
\showDOI{\tempurl}


\bibitem[\protect\citeauthoryear{Acquisti, Brandimarte, and
  Loewenstein}{Acquisti et~al\mbox{.}}{2015}]%
        {acquisti_behavior_2015}
\bibfield{author}{\bibinfo{person}{Alessandro Acquisti}, \bibinfo{person}{Laura
  Brandimarte}, {and} \bibinfo{person}{George Loewenstein}.}
  \bibinfo{year}{2015}\natexlab{}.
\newblock \showarticletitle{{Privacy and human behavior in the age of
  information}}.
\newblock \bibinfo{journal}{\emph{Science}} \bibinfo{volume}{347},
  \bibinfo{number}{6221} (\bibinfo{date}{Jan.} \bibinfo{year}{2015}),
  \bibinfo{pages}{509--514}.
\newblock
\showISSN{1095-9203}
\urldef\tempurl%
\url{https://doi.org/10.1126/science.aaa1465}
\showDOI{\tempurl}


\bibitem[\protect\citeauthoryear{{Alexa Internet, Inc.}}{{Alexa Internet,
  Inc.}}{2019}]%
        {alexa}
\bibfield{author}{\bibinfo{person}{{Alexa Internet, Inc.}}}
  \bibinfo{year}{2019}\natexlab{}.
\newblock \bibinfo{title}{{The top 500 sites on the Web}}.
\newblock
\newblock
\urldef\tempurl%
\url{https://www.alexa.com/topsites}
\showURL{%
\tempurl}


\bibitem[\protect\citeauthoryear{{Article 29 Data Protection Working
  Party}}{{Article 29 Data Protection Working Party}}{2016}]%
        {article_29_wp_cookie_sweep_2015}
\bibfield{author}{\bibinfo{person}{{Article 29 Data Protection Working
  Party}}.} \bibinfo{year}{2016}\natexlab{}.
\newblock \bibinfo{booktitle}{\emph{{Cookie Sweep Combined Analysis --
  Report}}}.
\newblock \bibinfo{type}{{T}echnical {R}eport} 14/EN WP 229.
  \bibinfo{institution}{European Commission}, \bibinfo{address}{Brussels,
  Belgium}.
\newblock


\bibitem[\protect\citeauthoryear{{Article 29 Data Protection Working
  Party}}{{Article 29 Data Protection Working Party}}{2018}]%
        {article_29_gdpr_consent}
\bibfield{author}{\bibinfo{person}{{Article 29 Data Protection Working
  Party}}.} \bibinfo{year}{2018}\natexlab{}.
\newblock \bibinfo{booktitle}{\emph{{Guidelines on consent under Regulation
  2016/679}}}.
\newblock \bibinfo{type}{{T}echnical {R}eport} 17/EN WP259 rev.01.
  \bibinfo{institution}{European Commission}.
\newblock


\bibitem[\protect\citeauthoryear{Boerman, Kruikemeier, and {Zuiderveen
  Borgesius}}{Boerman et~al\mbox{.}}{2018}]%
        {boerman_behavior_2018}
\bibfield{author}{\bibinfo{person}{Sophie~C. Boerman}, \bibinfo{person}{Sanne
  Kruikemeier}, {and} \bibinfo{person}{Frederik~J. {Zuiderveen Borgesius}}.}
  \bibinfo{year}{2018}\natexlab{}.
\newblock \showarticletitle{{Exploring Motivations for Online Privacy
  Protection Behavior: Insights From Panel Data}}.
\newblock \bibinfo{journal}{\emph{Communication Research}} \bibinfo{volume}{0},
  \bibinfo{number}{0} (\bibinfo{year}{2018}), \bibinfo{pages}{1--25}.
\newblock
\urldef\tempurl%
\url{https://doi.org/10.1177/0093650218800915}
\showDOI{\tempurl}


\bibitem[\protect\citeauthoryear{Burgess}{Burgess}{2018}]%
        {burgess_popups}
\bibfield{author}{\bibinfo{person}{Matt Burgess}.}
  \bibinfo{year}{2018}\natexlab{}.
\newblock \bibinfo{title}{The tyranny of GDPR popups and the websites failing
  to adapt}.
\newblock
\newblock
\urldef\tempurl%
\url{https://www.wired.co.uk/article/gdpr-cookies-eprivacy-regulation-popups}
\showURL{%
Retrieved April 22, 2019 from \tempurl}


\bibitem[\protect\citeauthoryear{Cantoni, Porta, Ricotti, and Zanin}{Cantoni
  et~al\mbox{.}}{2013}]%
        {cantoni_positioning_2013}
\bibfield{author}{\bibinfo{person}{Virginio Cantoni}, \bibinfo{person}{Marco
  Porta}, \bibinfo{person}{Stefania Ricotti}, {and} \bibinfo{person}{Francesca
  Zanin}.} \bibinfo{year}{2013}\natexlab{}.
\newblock \showarticletitle{{Banner positioning in the masthead area of online
  newspapers: an eye tracking study}}. In \bibinfo{booktitle}{\emph{14th
  International Conference on Computer Systems and Technologies}}
  \emph{(\bibinfo{series}{CompSysTech '13})}. \bibinfo{publisher}{ACM},
  \bibinfo{address}{New York, NY, USA}, \bibinfo{pages}{145--152}.
\newblock
\urldef\tempurl%
\url{https://doi.org/10.1145/2516775.2516789}
\showDOI{\tempurl}


\bibitem[\protect\citeauthoryear{Council)}{Council)}{2018}]%
        {forbrukerradet_deceived_2018}
\bibfield{author}{\bibinfo{person}{Forbrukerr{\aa}det (Norwegian~Consumer
  Council)}.} \bibinfo{year}{2018}\natexlab{}.
\newblock \bibinfo{booktitle}{\emph{{Deceived by Design -- How tech companies
  use dark patterns to discourage us from exercising our rights to privacy}}}.
\newblock \bibinfo{type}{{T}echnical {R}eport}. \bibinfo{address}{Oslo,
  Norway}.
\newblock


\bibitem[\protect\citeauthoryear{de~l'Informatique et des Libert\'{e}s
  (National Commission~on Informatics and Liberty)}{de~l'Informatique et des
  Libert\'{e}s (National Commission~on Informatics and Liberty)}{2018}]%
        {cnil_consent_2018}
\bibfield{author}{\bibinfo{person}{Commission~Nationale de~l'Informatique et
  des Libert\'{e}s (National Commission~on Informatics} {and}
  \bibinfo{person}{Liberty)}.} \bibinfo{year}{2018}\natexlab{}.
\newblock \bibinfo{title}{D\'{e}cision n\textsuperscript{o} MED 2018-042 du 30
  octobre 2018 mettant en demeure la soci\'{e}t\'{e} VECTAURY (Decision No. MED
  2018-042 of 30 October 2018 giving notice to the company VECTAURY)}.
\newblock
\newblock
\urldef\tempurl%
\url{https://www.legifrance.gouv.fr/affichCnil.do?id=CNILTEXT000037594451}
\showURL{%
Retrieved February 18, 2019 from \tempurl}


\bibitem[\protect\citeauthoryear{Degeling, Utz, Lentzsch, Hosseini, Schaub, and
  Holz}{Degeling et~al\mbox{.}}{2019}]%
        {degeling_gdpr_2019}
\bibfield{author}{\bibinfo{person}{Martin Degeling}, \bibinfo{person}{Christine
  Utz}, \bibinfo{person}{Christopher Lentzsch}, \bibinfo{person}{Henry
  Hosseini}, \bibinfo{person}{Florian Schaub}, {and} \bibinfo{person}{Thorsten
  Holz}.} \bibinfo{year}{2019}\natexlab{}.
\newblock \showarticletitle{{We Value Your Privacy ... Now Take Some Cookies:
  Measuring the GDPR's Impact on Web Privacy}}. In
  \bibinfo{booktitle}{\emph{26th Annual Network and Distributed System Security
  Symposium}} \emph{(\bibinfo{series}{NDSS '19})}. \bibinfo{publisher}{Internet
  Society}.
\newblock


\bibitem[\protect\citeauthoryear{Egelman, Cranor, and Hong}{Egelman
  et~al\mbox{.}}{2008}]%
        {egelman_phishing_2008}
\bibfield{author}{\bibinfo{person}{Serge Egelman},
  \bibinfo{person}{Lorrie~Faith Cranor}, {and} \bibinfo{person}{Jason Hong}.}
  \bibinfo{year}{2008}\natexlab{}.
\newblock \showarticletitle{{You've Been Warned: An Empirical Study of the
  Effectiveness of Web Browser Phishing Warnings}}. In
  \bibinfo{booktitle}{\emph{Conference on Human Factors in Computing Systems}}
  \emph{(\bibinfo{series}{CHI '08})}. \bibinfo{publisher}{ACM},
  \bibinfo{address}{New York, NY, USA}, \bibinfo{pages}{1065--1074}.
\newblock
\showISBNx{978-1-60558-011-1}
\urldef\tempurl%
\url{https://doi.org/10.1145/1357054.1357219}
\showDOI{\tempurl}


\bibitem[\protect\citeauthoryear{Europe}{Europe}{2019}]%
        {iab_framework_2019}
\bibfield{author}{\bibinfo{person}{Interactive Advertising~Bureau Europe}.}
  \bibinfo{year}{2019}\natexlab{}.
\newblock \bibinfo{title}{{GDPR Transparency and Consent Framework}}.
\newblock
  \bibinfo{howpublished}{\url{https://iabtechlab.com/standards/gdpr-transparency-and-consent-framework/}}.
\newblock
\newblock
\shownote{[Online; accessed 2 May 2019].}


\bibitem[\protect\citeauthoryear{{European Data Protection Board}}{{European
  Data Protection Board}}{2019}]%
        {edpb_gdpr_eprivacy_2019}
\bibfield{author}{\bibinfo{person}{{European Data Protection Board}}.}
  \bibinfo{year}{2019}\natexlab{}.
\newblock \bibinfo{booktitle}{\emph{{Opinion 5/2019 on the interplay between
  the ePrivacy Directive and the GDPR, in particular regarding the competence,
  tasks and powers of data protection authorities}}}.
\newblock \bibinfo{type}{{T}echnical {R}eport} 5/2019.
\newblock


\bibitem[\protect\citeauthoryear{Felt, Ainslie, Reeder, Consolvo, Thyagaraja,
  Bettes, and Grimes}{Felt et~al\mbox{.}}{2015}]%
        {felt_ssl_2015}
\bibfield{author}{\bibinfo{person}{Adrienne~Porter Felt}, \bibinfo{person}{Alex
  Ainslie}, \bibinfo{person}{Robert~W. Reeder}, \bibinfo{person}{Sunny
  Consolvo}, \bibinfo{person}{Somas Thyagaraja}, \bibinfo{person}{Helen Bettes,
  Alan ad~Harris}, {and} \bibinfo{person}{Jeff Grimes}.}
  \bibinfo{year}{2015}\natexlab{}.
\newblock \showarticletitle{{Improving SSL Warnings: Comprehension and
  Adherence}}. In \bibinfo{booktitle}{\emph{{33rd Annual ACM Conference on
  Human Factors in Computing Systems}}} \emph{(\bibinfo{series}{CHI '15})}.
  \bibinfo{publisher}{ACM}, \bibinfo{address}{New York, NY, USA},
  \bibinfo{pages}{2893--2902}.
\newblock
\showISBNx{978-1-4503-3145-6}
\urldef\tempurl%
\url{https://doi.org/10.1145/2702123.2702442}
\showDOI{\tempurl}


\bibitem[\protect\citeauthoryear{Friedman}{Friedman}{2019}]%
        {friedman_privacyux_2019}
\bibfield{author}{\bibinfo{person}{Vitaly Friedman}.}
  \bibinfo{year}{2019}\natexlab{}.
\newblock \bibinfo{title}{Privacy UX: Better Cookie Consent Experiences}.
\newblock
\newblock
\urldef\tempurl%
\url{https://www.smashingmagazine.com/2019/04/privacy-ux-better-cookie-consent-experiences/}
\showURL{%
Retrieved May 7, 2019 from \tempurl}


\bibitem[\protect\citeauthoryear{Garlach and Suthers}{Garlach and
  Suthers}{2018}]%
        {garlach_adchoices_2018}
\bibfield{author}{\bibinfo{person}{Stacia Garlach} {and}
  \bibinfo{person}{Daniel Suthers}.} \bibinfo{year}{2018}\natexlab{}.
\newblock \showarticletitle{{`I'm supposed to see that?' AdChoices Usability in
  the Mobile Environment}}. In \bibinfo{booktitle}{\emph{Hawaii International
  Conference on System Sciences}}. \bibinfo{publisher}{University of Hawai`i at
  M\={a}noa}, \bibinfo{address}{Honolulu, HI, USA},
  \bibinfo{pages}{3779--3788}.
\newblock
\showISBNx{978-0-9981331-1-9}
\urldef\tempurl%
\url{https://doi.org/10.24251/hicss.2018.476}
\showDOI{\tempurl}


\bibitem[\protect\citeauthoryear{Ha, Inkpen, Al~Shaar, and Hdeib}{Ha
  et~al\mbox{.}}{2006}]%
        {ha_cookie_perception_2006}
\bibfield{author}{\bibinfo{person}{Vicki Ha}, \bibinfo{person}{Kori Inkpen},
  \bibinfo{person}{Farah Al~Shaar}, {and} \bibinfo{person}{Lina Hdeib}.}
  \bibinfo{year}{2006}\natexlab{}.
\newblock \showarticletitle{An Examination of User Perception and Misconception
  of Internet Cookies}. In \bibinfo{booktitle}{\emph{CHI '06 Extended Abstracts
  on Human Factors in Computing Systems}} \emph{(\bibinfo{series}{CHI EA
  '06})}. \bibinfo{publisher}{ACM}, \bibinfo{address}{New York, NY, USA},
  \bibinfo{pages}{833--838}.
\newblock
\showISBNx{1-59593-298-4}
\urldef\tempurl%
\url{https://doi.org/10.1145/1125451.1125615}
\showDOI{\tempurl}


\bibitem[\protect\citeauthoryear{Habib, Zou, Jannu, Sridhar, Swoopes, Acquisti,
  Cranor, Sadeh, and Schaub}{Habib et~al\mbox{.}}{2019}]%
        {habib_optout_2019}
\bibfield{author}{\bibinfo{person}{Hana Habib}, \bibinfo{person}{Yixin Zou},
  \bibinfo{person}{Aditi Jannu}, \bibinfo{person}{Neha Sridhar},
  \bibinfo{person}{Chelse Swoopes}, \bibinfo{person}{Alessandro Acquisti},
  \bibinfo{person}{Lorrie~Faith Cranor}, \bibinfo{person}{Norman Sadeh}, {and}
  \bibinfo{person}{Florian Schaub}.} \bibinfo{year}{2019}\natexlab{}.
\newblock \showarticletitle{{An Empirical Analysis of Data Deletion and Opt-Out
  Choices on 150 Websites}}. In \bibinfo{booktitle}{\emph{Fifteenth Symposium
  On Usable Privacy and Security}} \emph{(\bibinfo{series}{SOUPS 2019})}.
  \bibinfo{publisher}{{USENIX} Association}, \bibinfo{pages}{387--406}.
\newblock
\urldef\tempurl%
\url{https://www.usenix.org/conference/soups2019/presentation/habib}
\showURL{%
\tempurl}


\bibitem[\protect\citeauthoryear{Kladnik}{Kladnik}{2019}]%
        {i_dont_care_about_cookies_2019}
\bibfield{author}{\bibinfo{person}{Daniel Kladnik}.}
  \bibinfo{year}{2019}\natexlab{}.
\newblock \bibinfo{title}{{I don't care about cookies 3.0.0}}.
\newblock
  \bibinfo{howpublished}{\url{https://www.i-dont-care-about-cookies.eu/}}.
\newblock
\newblock
\shownote{[Online; accessed 2 May 2019].}


\bibitem[\protect\citeauthoryear{Kulyk, Hilt, Gerber, and Volkamer}{Kulyk
  et~al\mbox{.}}{2018a}]%
        {kulyk_cookie_perceptions_2018}
\bibfield{author}{\bibinfo{person}{Oksana Kulyk}, \bibinfo{person}{Annika
  Hilt}, \bibinfo{person}{Nina Gerber}, {and} \bibinfo{person}{Melanie
  Volkamer}.} \bibinfo{year}{2018}\natexlab{a}.
\newblock \showarticletitle{{``This Website Uses Cookies'': Users' Perceptions
  and Reactions to the Cookie Disclaimer}}. In \bibinfo{booktitle}{\emph{3rd
  European Workshop on Usable Security}} \emph{(\bibinfo{series}{EuroUSec
  2018})}. \bibinfo{address}{London, England}, 11.
\newblock


\bibitem[\protect\citeauthoryear{Kulyk, Mayer, K{\"a}fer, and Volkamer}{Kulyk
  et~al\mbox{.}}{2018b}]%
        {kulyk_cookie_interface_2018}
\bibfield{author}{\bibinfo{person}{Oksana Kulyk}, \bibinfo{person}{Peter
  Mayer}, \bibinfo{person}{Oliver K{\"a}fer}, {and} \bibinfo{person}{Melanie
  Volkamer}.} \bibinfo{year}{2018}\natexlab{b}.
\newblock \showarticletitle{{A Concept and Evaluation of Usable and
  Fine-Grained Privacy-Friendly Cookie Settings Interface}}. In
  \bibinfo{booktitle}{\emph{{17th IEEE International Conference On Trust,
  Security And Privacy In Computing And Communications}}}
  \emph{(\bibinfo{series}{TrustCom 2018})}. \bibinfo{publisher}{IEEE},
  \bibinfo{address}{Piscataway, NJ, USA}.
\newblock


\bibitem[\protect\citeauthoryear{Leon, Ur, Shay, Wang, Balebako, and
  Cranor}{Leon et~al\mbox{.}}{2012}]%
        {leon_optout_2012}
\bibfield{author}{\bibinfo{person}{Pedro Leon}, \bibinfo{person}{Blase Ur},
  \bibinfo{person}{Richard Shay}, \bibinfo{person}{Yang Wang},
  \bibinfo{person}{Rebecca Balebako}, {and} \bibinfo{person}{Lorrie Cranor}.}
  \bibinfo{year}{2012}\natexlab{}.
\newblock \showarticletitle{{Why Johnny can't opt out: a usability evaluation
  of tools to limit online behavioral advertising}}. In
  \bibinfo{booktitle}{\emph{Conference on Human Factors in Computing Systems}}
  \emph{(\bibinfo{series}{CHI '12})}. \bibinfo{publisher}{ACM},
  \bibinfo{address}{New York, NY, USA}, \bibinfo{pages}{589--598}.
\newblock
\showISBNx{978-1-4503-1015-4}
\urldef\tempurl%
\url{https://doi.org/10.1145/2207676.2207759}
\showDOI{\tempurl}


\bibitem[\protect\citeauthoryear{Liu, White, and Dumais}{Liu
  et~al\mbox{.}}{2010}]%
        {liu_dwell_2010}
\bibfield{author}{\bibinfo{person}{Chao Liu}, \bibinfo{person}{Ryen~W. White},
  {and} \bibinfo{person}{Susan Dumais}.} \bibinfo{year}{2010}\natexlab{}.
\newblock \showarticletitle{{Understanding Web Browsing Behaviors Through
  Weibull Analysis of Dwell Time}}. In \bibinfo{booktitle}{\emph{{33rd
  International ACM SIGIR Conference on Research and Development in Information
  Retrieval}}} \emph{(\bibinfo{series}{SIGIR '10})}. \bibinfo{publisher}{ACM},
  \bibinfo{address}{New York, NY, USA}, \bibinfo{pages}{379--386}.
\newblock
\showISBNx{978-1-4503-0153-4}
\urldef\tempurl%
\url{https://doi.org/10.1145/1835449.1835513}
\showDOI{\tempurl}


\bibitem[\protect\citeauthoryear{Manafactory}{Manafactory}{2019}]%
        {manafactory_ginger_2018}
\bibfield{author}{\bibinfo{person}{Manafactory}.}
  \bibinfo{year}{2019}\natexlab{}.
\newblock \bibinfo{title}{{Ginger -- EU Cookie Law}}.
\newblock \bibinfo{howpublished}{\url{https://wordpress.org/plugins/ginger/}}.
\newblock
\newblock
\shownote{[Online; accessed 22 August 2019].}


\bibitem[\protect\citeauthoryear{Martin}{Martin}{2016}]%
        {martin_privacy_2016}
\bibfield{author}{\bibinfo{person}{Kirsten Martin}.}
  \bibinfo{year}{2016}\natexlab{}.
\newblock \showarticletitle{{Do Privacy Notices Matter? Comparing the Impact of
  Violating Formal Privacy Notices and Informal Privacy Norms on Consumer Trust
  Online}}.
\newblock \bibinfo{journal}{\emph{The Journal of Legal Studies}}
  \bibinfo{volume}{45}, \bibinfo{number}{S2} (\bibinfo{date}{June}
  \bibinfo{year}{2016}), \bibinfo{pages}{S191--S215}.
\newblock
\showISSN{1537-5366}
\urldef\tempurl%
\url{https://doi.org/10.1086/688488}
\showDOI{\tempurl}


\bibitem[\protect\citeauthoryear{Mathur, Acar, Friedman, Lucherini, Mayer, and
  Chetty}{Mathur et~al\mbox{.}}{2019}]%
        {mathur_patterns_2019}
\bibfield{author}{\bibinfo{person}{Arunesh Mathur}, \bibinfo{person}{Gunes
  Acar}, \bibinfo{person}{Michael Friedman}, \bibinfo{person}{Elena Lucherini},
  \bibinfo{person}{Jonathan Mayer}, {and} \bibinfo{person}{Marsh Chetty}.}
  \bibinfo{year}{2019}\natexlab{}.
\newblock \showarticletitle{{Dark Patterns at Scale: Findings from a Crawl of
  11K Shopping Websites}}.
\newblock  (\bibinfo{year}{2019}).
\newblock
\showeprint{1907.07032}


\bibitem[\protect\citeauthoryear{Mayer and Mitchell}{Mayer and
  Mitchell}{2012}]%
        {mayer_DNT_2012}
\bibfield{author}{\bibinfo{person}{Jonathan~R. Mayer} {and}
  \bibinfo{person}{John~C. Mitchell}.} \bibinfo{year}{2012}\natexlab{}.
\newblock \showarticletitle{{Third-Party Web Tracking: Policy and Technology}}.
  In \bibinfo{booktitle}{\emph{2012 IEEE Symposium on Security and Privacy}}
  \emph{(\bibinfo{series}{SP '12})}. \bibinfo{publisher}{IEEE Computer
  Society}, \bibinfo{address}{Washington, DC, USA}, \bibinfo{pages}{413--427}.
\newblock
\showISBNx{978-0-7695-4681-0}
\urldef\tempurl%
\url{https://doi.org/10.1109/SP.2012.47}
\showDOI{\tempurl}


\bibitem[\protect\citeauthoryear{McDonald and Cranor}{McDonald and
  Cranor}{2010}]%
        {mcdonald_cookies_2010}
\bibfield{author}{\bibinfo{person}{Aleecia~M. McDonald} {and}
  \bibinfo{person}{Lorrie~Faith Cranor}.} \bibinfo{year}{2010}\natexlab{}.
\newblock \showarticletitle{{Americans' Attitudes About Internet Behavioral
  Advertising Practices}}. In \bibinfo{booktitle}{\emph{9th Annual ACM Workshop
  on Privacy in the Electronic Society}} \emph{(\bibinfo{series}{WPES '10})}.
  \bibinfo{publisher}{ACM}, \bibinfo{address}{New York, NY, USA},
  \bibinfo{pages}{63--72}.
\newblock
\showISBNx{978-1-4503-0096-4}
\urldef\tempurl%
\url{https://doi.org/10.1145/1866919.1866929}
\showDOI{\tempurl}


\bibitem[\protect\citeauthoryear{O'Neill}{O'Neill}{2018}]%
        {oneill_dnt_gdpr_2018}
\bibfield{author}{\bibinfo{person}{Mike O'Neill}.}
  \bibinfo{year}{2018}\natexlab{}.
\newblock \bibinfo{title}{{Do Not Track and the GDPR}}.
\newblock
\newblock
\urldef\tempurl%
\url{https://www.w3.org/blog/2018/06/do-not-track-and-the-gdpr/}
\showURL{%
Retrieved May 15, 2019 from \tempurl}


\bibitem[\protect\citeauthoryear{Rao, Schaub, Sadeh, Acquisti, and Kang}{Rao
  et~al\mbox{.}}{2016}]%
        {rao_mismatched_2016}
\bibfield{author}{\bibinfo{person}{Ashwini Rao}, \bibinfo{person}{Florian
  Schaub}, \bibinfo{person}{Norman Sadeh}, \bibinfo{person}{Alessandro
  Acquisti}, {and} \bibinfo{person}{Ruogo Kang}.}
  \bibinfo{year}{2016}\natexlab{}.
\newblock \showarticletitle{{Expecting the Unexpected: Understanding Mismatched
  Privacy Expectations Online}}. In \bibinfo{booktitle}{\emph{Twelfth Symposium
  On Usable Privacy and Security}} \emph{(\bibinfo{series}{SOUPS '16})}.
  \bibinfo{publisher}{{USENIX} Association}, \bibinfo{pages}{77--96}.
\newblock
\showISBNx{978-1-931971-31-7}
\urldef\tempurl%
\url{https://www.usenix.org/conference/soups2016/technical-sessions/presentation/rao}
\showURL{%
\tempurl}


\bibitem[\protect\citeauthoryear{Reeder, Felt, Consolvo, Malkin, Thompson, and
  Egelman}{Reeder et~al\mbox{.}}{2018}]%
        {reeder_browser_warnings_2018}
\bibfield{author}{\bibinfo{person}{Robert~W. Reeder},
  \bibinfo{person}{Adrienne~Porter Felt}, \bibinfo{person}{Sunny Consolvo},
  \bibinfo{person}{Nathan Malkin}, \bibinfo{person}{Christopher Thompson},
  {and} \bibinfo{person}{Serge Egelman}.} \bibinfo{year}{2018}\natexlab{}.
\newblock \showarticletitle{{An Experience Sampling Study of User Reactions to
  Browser Warnings in the Field}}. In \bibinfo{booktitle}{\emph{Conference on
  Human Factors in Computing Systems}} \emph{(\bibinfo{series}{CHI '18})}.
  \bibinfo{publisher}{ACM}, \bibinfo{address}{New York, NY, USA}.
\newblock
\showISBNx{978-1-4503-5620-6}
\urldef\tempurl%
\url{https://doi.org/10.1145/3173574.3174086}
\showDOI{\tempurl}


\bibitem[\protect\citeauthoryear{Ryan}{Ryan}{2017a}]%
        {ryan_consent_2017}
\bibfield{author}{\bibinfo{person}{Johnny Ryan}.}
  \bibinfo{year}{2017}\natexlab{a}.
\newblock \bibinfo{title}{Research result: what percentage will consent to
  tracking for...}
\newblock
\newblock
\urldef\tempurl%
\url{https://pagefair.com/blog/2017/new-research-how-many-consent-to-tracking/}
\showURL{%
\tempurl}


\bibitem[\protect\citeauthoryear{Ryan}{Ryan}{2017b}]%
        {pagefairadblock}
\bibfield{author}{\bibinfo{person}{Johnny Ryan}.}
  \bibinfo{year}{2017}\natexlab{b}.
\newblock \bibinfo{booktitle}{\emph{{The state of the blocked web -- 2017
  Global Adblock Report}}}.
\newblock \bibinfo{type}{{T}echnical {R}eport}.
  \bibinfo{institution}{PageFair}.
\newblock
\urldef\tempurl%
\url{https://pagefair.com/downloads/2017/01/PageFair-2017-Adblock-Report.pdf}
\showURL{%
Retrieved May 8, 2019 from \tempurl}


\bibitem[\protect\citeauthoryear{Ryan}{Ryan}{2018}]%
        {ryan_iab_2018}
\bibfield{author}{\bibinfo{person}{Johnny Ryan}.}
  \bibinfo{year}{2018}\natexlab{}.
\newblock \bibinfo{title}{French regulator shows deep flaws in IAB's consent
  framework and RTB}.
\newblock
\newblock
\urldef\tempurl%
\url{https://brave.com/cnil-consent-rtb/}
\showURL{%
Retrieved May 8, 2019 from \tempurl}


\bibitem[\protect\citeauthoryear{Ryan}{Ryan}{2019}]%
        {ryan_complaint_2019}
\bibfield{author}{\bibinfo{person}{Johnny Ryan}.}
  \bibinfo{year}{2019}\natexlab{}.
\newblock \bibinfo{title}{{Formal GDPR complaint against IAB Europe`s ``cookie
  wall'' and GDPR consent guidance.}}
\newblock
\newblock
\urldef\tempurl%
\url{https://brave.com/iab-cookie-wall/}
\showURL{%
Retrieved May 10, 2019 from \tempurl}


\bibitem[\protect\citeauthoryear{Sanchez-Rola, Dell'Amico, Kotzias, Balzarotti,
  Bilge, Vervier, and Santos}{Sanchez-Rola et~al\mbox{.}}{2019}]%
        {sanchez-rola_optout_2019}
\bibfield{author}{\bibinfo{person}{Iskander Sanchez-Rola},
  \bibinfo{person}{Matteo Dell'Amico}, \bibinfo{person}{Platon Kotzias},
  \bibinfo{person}{Davide Balzarotti}, \bibinfo{person}{Leyla Bilge},
  \bibinfo{person}{Pierre-Antoine Vervier}, {and} \bibinfo{person}{Igor
  Santos}.} \bibinfo{year}{2019}\natexlab{}.
\newblock \showarticletitle{{Can I Opt Out Yet? GDPR and the Global Illusion of
  Cookie Control}}. In \bibinfo{booktitle}{\emph{ACM ASIA Conference on
  Computer and Communications Security}} \emph{(\bibinfo{series}{AsiaCCS
  '19})}. \bibinfo{publisher}{ACM}, \bibinfo{address}{New York, NY, USA}.
\newblock
\urldef\tempurl%
\url{https://doi.org/10.1145/3321705.3329806}
\showDOI{\tempurl}


\bibitem[\protect\citeauthoryear{Schaub, Balebako, Durity, and Cranor}{Schaub
  et~al\mbox{.}}{2015}]%
        {schaub_design_2015}
\bibfield{author}{\bibinfo{person}{Florian Schaub}, \bibinfo{person}{Rebecca
  Balebako}, \bibinfo{person}{Adam~L. Durity}, {and}
  \bibinfo{person}{Lorrie~Faith Cranor}.} \bibinfo{year}{2015}\natexlab{}.
\newblock \showarticletitle{{A Design Space for Effective Privacy Notices}}. In
  \bibinfo{booktitle}{\emph{Eleventh Symposium On Usable Privacy and Security}}
  \emph{(\bibinfo{series}{SOUPS '15})}. \bibinfo{publisher}{The {USENIX}
  Association}, \bibinfo{address}{Ottawa}, \bibinfo{pages}{1--17}.
\newblock
\showISBNx{978-1-931971-249}
\urldef\tempurl%
\url{https://doi.org/10.1145/567752.567774}
\showDOI{\tempurl}


\bibitem[\protect\citeauthoryear{Silic}{Silic}{2016}]%
        {silic_colour_2016}
\bibfield{author}{\bibinfo{person}{Mario Silic}.}
  \bibinfo{year}{2016}\natexlab{}.
\newblock \showarticletitle{{Understanding Colour Impact on Warning Messages:
  Evidence from US and India}}. In \bibinfo{booktitle}{\emph{2016 CHI
  Conference Extended Abstracts on Human Factors in Computing Systems}}
  \emph{(\bibinfo{series}{CHI EA '16})}. \bibinfo{publisher}{ACM},
  \bibinfo{address}{New York, NY, USA}, \bibinfo{pages}{2954--2960}.
\newblock
\urldef\tempurl%
\url{https://doi.org/10.1145/2851581.2892276}
\showDOI{\tempurl}


\bibitem[\protect\citeauthoryear{S{\o}rensen and Kosta}{S{\o}rensen and
  Kosta}{2019}]%
        {sorensen_gdpr_tracking_2019}
\bibfield{author}{\bibinfo{person}{Jannick S{\o}rensen} {and}
  \bibinfo{person}{Sokol Kosta}.} \bibinfo{year}{2019}\natexlab{}.
\newblock \showarticletitle{{Before and After GDPR: The Changes in Third Party
  Presence at Public and Private European Websites}}. In
  \bibinfo{booktitle}{\emph{The 2019 World Wide Web Conference}}
  \emph{(\bibinfo{series}{WWW '19})}. \bibinfo{publisher}{ACM},
  \bibinfo{address}{New York, NY, USA}, \bibinfo{pages}{1590--1600}.
\newblock
\urldef\tempurl%
\url{https://doi.org/10.1145/3308558.3313524}
\showDOI{\tempurl}


\bibitem[\protect\citeauthoryear{{State of California Legislative
  Counsel}}{{State of California Legislative Counsel}}{2018}]%
        {ccpa_2018}
\bibfield{author}{\bibinfo{person}{{State of California Legislative Counsel}}.}
  \bibinfo{year}{2018}\natexlab{}.
\newblock \bibinfo{title}{{Assembly Bill No. 375 -- Chapter 55}}.
\newblock
\newblock


\bibitem[\protect\citeauthoryear{Thaler and Sunstein}{Thaler and
  Sunstein}{2009}]%
        {thaler_nudge_2009}
\bibfield{author}{\bibinfo{person}{Richard~H. Thaler} {and}
  \bibinfo{person}{Cass~R. Sunstein}.} \bibinfo{year}{2009}\natexlab{}.
\newblock \bibinfo{booktitle}{\emph{Nudge: Improving Decisions About Health,
  Wealth, and Happiness}}.
\newblock \bibinfo{publisher}{Penguin Books}, \bibinfo{address}{New York, NY,
  USA}.
\newblock
\showISBNx{978-0143115267}


\bibitem[\protect\citeauthoryear{{The European Parliament and the Council of
  the European Union}}{{The European Parliament and the Council of the European
  Union}}{2002}]%
        {eprivacy_directive_2002}
\bibfield{author}{\bibinfo{person}{{The European Parliament and the Council of
  the European Union}}.} \bibinfo{year}{2002}\natexlab{}.
\newblock \bibinfo{title}{{Directive 2002/58/EC of the European Parliament and
  of the Council of 12 July 2002 concerning the processing of personal data and
  the protection of privacy in the electronic communications sector}}.
\newblock \bibinfo{howpublished}{Official Journal of the European Communities}.
\newblock


\bibitem[\protect\citeauthoryear{{The European Parliament and the Council of
  the European Union}}{{The European Parliament and the Council of the European
  Union}}{2009}]%
        {cookie_directive_2009}
\bibfield{author}{\bibinfo{person}{{The European Parliament and the Council of
  the European Union}}.} \bibinfo{year}{2009}\natexlab{}.
\newblock \bibinfo{title}{{Directive 2009/136/EC of the European Parliament and
  of the Council of 25 November 2009 amending Directive 2002/22/EC, Directive
  2002/58/EC and Regulation (EC) No 2006/2004}}.
\newblock \bibinfo{howpublished}{Official Journal of the European Union, L
  337/11}.
\newblock


\bibitem[\protect\citeauthoryear{{The European Parliament and the Council of
  the European Union}}{{The European Parliament and the Council of the European
  Union}}{2016}]%
        {gdpr_2016}
\bibfield{author}{\bibinfo{person}{{The European Parliament and the Council of
  the European Union}}.} \bibinfo{year}{2016}\natexlab{}.
\newblock \bibinfo{title}{{Regulation (EU) 2016/679 of the European Parliament
  and of the Council of 27 April 2016 on the protection of natural persons with
  regard to the processing of personal data and on the free movement of such
  data, and repealing Directive 95/46/EC (General Data Protection
  Regulation)}}.
\newblock \bibinfo{howpublished}{Official Journal of the European Union, L
  119/1}.
\newblock


\bibitem[\protect\citeauthoryear{Turow, Hennessy, and Draper}{Turow
  et~al\mbox{.}}{2018}]%
        {turow_persistent_2018}
\bibfield{author}{\bibinfo{person}{Joseph Turow}, \bibinfo{person}{Michael
  Hennessy}, {and} \bibinfo{person}{Nora Draper}.}
  \bibinfo{year}{2018}\natexlab{}.
\newblock \showarticletitle{{Persistent Misperceptions: Americans' Misplaced
  Confidence in Privacy Policies, 2003--2015}}.
\newblock \bibinfo{journal}{\emph{Journal of Broadcasting \& Electronic Media}}
  \bibinfo{volume}{62}, \bibinfo{number}{3} (\bibinfo{date}{July}
  \bibinfo{year}{2018}), \bibinfo{pages}{461--478}.
\newblock
\showISSN{1550-6878}
\urldef\tempurl%
\url{https://doi.org/10.1080/08838151.2018.1451867}
\showDOI{\tempurl}


\bibitem[\protect\citeauthoryear{Urban, Degeling, Holz, and Pohlmann}{Urban
  et~al\mbox{.}}{2019}]%
        {urban_perspectives_2019}
\bibfield{author}{\bibinfo{person}{Tobias Urban}, \bibinfo{person}{Martin
  Degeling}, \bibinfo{person}{Thorsten Holz}, {and} \bibinfo{person}{Norbert
  Pohlmann}.} \bibinfo{year}{2019}\natexlab{}.
\newblock \showarticletitle{Perspectives on {Transparency} {Tools} for {Online}
  {Advertising}}. In \bibinfo{booktitle}{\emph{35th {Annual} {Computer}
  {Security} {Applications} {Conference} ({ACSAC})}}. \bibinfo{publisher}{ACM},
  \bibinfo{address}{San Juan}, 14.
\newblock


\bibitem[\protect\citeauthoryear{{van Eijk}, Asghari, Winter, and
  Narayanan}{{van Eijk} et~al\mbox{.}}{2019}]%
        {vaneijk_location_2019}
\bibfield{author}{\bibinfo{person}{Rob {van Eijk}}, \bibinfo{person}{Hadi
  Asghari}, \bibinfo{person}{Philipp Winter}, {and} \bibinfo{person}{Arvind
  Narayanan}.} \bibinfo{year}{2019}\natexlab{}.
\newblock \showarticletitle{{The Impact of User Location on Cookie Notices
  (Inside and Outside of the European Union)}}. In
  \bibinfo{booktitle}{\emph{Workshop on Technology and Consumer Protection}}
  \emph{(\bibinfo{series}{ConPro '19})}. \bibinfo{publisher}{IEEE}.
\newblock


\bibitem[\protect\citeauthoryear{Weinmann, Schneider, and {vom
  Brocke}}{Weinmann et~al\mbox{.}}{2016}]%
        {weinmann_nudging_2016}
\bibfield{author}{\bibinfo{person}{Markus Weinmann}, \bibinfo{person}{Christoph
  Schneider}, {and} \bibinfo{person}{Jan {vom Brocke}}.}
  \bibinfo{year}{2016}\natexlab{}.
\newblock \showarticletitle{{Digital Nudging}}.
\newblock \bibinfo{journal}{\emph{Business \& Information Systems Engineering}}
  \bibinfo{volume}{58}, \bibinfo{number}{6} (\bibinfo{date}{Dec.}
  \bibinfo{year}{2016}), \bibinfo{pages}{433--436}.
\newblock
\showISSN{1867-0202}
\urldef\tempurl%
\url{https://doi.org/10.1007/s12599-016-0453-1}
\showDOI{\tempurl}


\end{thebibliography}

% 
% If your work has an appendix, this is the place to put it.
\appendix
\section{Timing in Experiment 2}
\label{sec:timing}

\begin{table}[bh]
    \centering
    \caption{Average time in seconds until users submitted decision in Experiment~2, if decision was made within the first three minutes}
    \label{tab:timingr2}

        \begin{tabular}{llllll}
            \toprule
            Banner  & Type & \# Users & Mean & Median & SD \\
            \midrule 
            
No Option & & 4174 & 6.51 & 4 & 15.55\\
Confirmation & non-nudging & 2984 & 10.65 & 5 & 51.25\\
 & nudging & 3634 & 9.11 & 4 & 37.78\\
Binary& non-nudging & 4134 & 15.36 & 4 & 72.47\\
& nudging & 4097 & 13.51 & 4 & 75.59\\
Category & non-nudging & 2523 & 17.93 & 8 & 87.16\\
 & nudging & 2798 & 13.98 & 7 & 64.01\\
Vendor & non-nudging & 2346 & 13.76 & 8 & 42.38\\
& nudging & 2741 & 21.18 & 7 & 115.26\\

            \bottomrule
    \end{tabular}
\end{table}

\onecolumn
\section{Survey and Responses}
\label{sec:survey-responses}

\textsuperscript{R} indicates answers displayed in random order. All questions and answers were translated from German as true to the original as possible.

\begin{table*}[h!tb]
	\centering
	\small
	\caption*{Motivation for Interacting With the Cookie Consent Notice}
	\label{tab:survey-motivation}
	\begin{threeparttable}
		\begin{tabular}{@{}lrrrrr}
			\toprule
			\multicolumn{6}{l}{Q1-clicked\textsuperscript{a}: You just clicked the cookie consent notice\textsuperscript{b} on the website [WEBSITE\_NAME]. Which of the following statements describe your}\\
			\multicolumn{6}{l}{ motivation to click the notice? I clicked the cookie consent notice ... [multiple choice]}\\
			\midrule
			& \textbf{Exp. 1} & \textbf{Exp. 2} & \textbf{Exp. 3} & \textbf{Total} & \textbf{\,\%} \\
			... to protect me from dangers from the Internet.\textsuperscript{R} & 0 & 3 & 3 & 6 &  9.8\,\%\\
			... to protect my privacy on the Internet.\textsuperscript{R} & 0 & 5 & 6 & 11 & 18.0\,\%\\
			... because the website does not work otherwise.\textsuperscript{R} & 2 & 11 & 3 & 16 & 26.2\,\%\\
			... to see less ads.\textsuperscript{R} & 1 & 1 & 3 & 5 & 8.2\,\%\\
			... out of habit.\textsuperscript{R} & 1 & 10 & 2 & 13 & 21.3\,\%\\
			... because the notice distracts me from viewing the website.\textsuperscript{R} & 6 & 25 & 13 & 44 & 72.1\,\%\\
			Other: [free text] & 0 & 0 & 1 & 1 & 1.6\,\%\\
			I do not know why I clicked the notice. & 1 & 1 & 1 & 3 & 4.9\,\%\\
			I prefer not to answer. & 0 & 0 & 0 & 0 & 0\,\%\\
			\textbf{\# Answers} & \textbf{11} & \textbf{56} & \textbf{32} & \textbf{99} & \textbf{}\\
			\textbf{\# Participants} & \textbf{8} & \textbf{34} & \textbf{19} & \textbf{61} & \textbf{}\\
			&&&&&\\
			\midrule
			\midrule
			\multicolumn{6}{l}{Q1-notclicked:\textsuperscript{a}  You did not click the cookie consent notice\textsuperscript{b} on the website [WEBSITE\_NAME]. Which of the following statements describe your}\\
			\multicolumn{6}{l}{ motivation to not click the notice? I did not click the cookie consent notice ... [multiple choice]}\\
			\midrule
			& \textbf{Exp. 1} & \textbf{Exp. 2} & \textbf{Exp. 3} & \textbf{Total} & \textbf{\,\%} \\
			
			... because I have not noticed it.\textsuperscript{R} & 4 & 11 & 5 & 20 & 40.8\,\%\\
	        ... because it did not offer enough choices.\textsuperscript{R} & 0 & 0 & 3 & 3 & 6.1\,\%\\
	        ... because I do not know what happens if I click the notice.\textsuperscript{R} & 1 & 6 & 4 & 11 & 22.4\,\%\\
	        ... because I think that my selection does not have any effect.\textsuperscript{R} & 1 & 4 & 4 & 9 & 18.4\,\% \\
	        ... because I do not know what cookies are.\textsuperscript{R} & 0 & 2 & 0 & 2 & 4.1\,\%\\
	        ... because I do not care which cookies the website uses.\textsuperscript{R}\tnote{c} & 1 & 3 & 2 & 6 & 12.2\,\% \\
	        ... Other: [free text] & 1 & 10 & 2 & 13 & 26.5\,\%\\
	        ... I do not know why I did not click the cookie consent notice. & 1 & 0 & 0 & 1 & 2.0\,\%\\
	        ... I prefer not to answer. & 0 & 2 & 0 & 2 & 4.1\,\%\\
			\textbf{\# Answers} & \textbf{9} & \textbf{38} & \textbf{20} & \textbf{67} & \textbf{}\\
			\textbf{\# Participants} & \textbf{8} & \textbf{26} & \textbf{15} & \textbf{49} & \textbf{}\\
			\midrule
			\bottomrule
		\end{tabular}
	\centering
	\begin{tablenotes}
		\item[a] Q1-clicked and Q1-notclicked were only displayed to participants who clicked / did not click the notice, respectively.
		\item[b] In Experiment~3, ``cookie consent notice'' was changed to ``privacy notice'' in the conditions Non-Technical--PP Link and Non-Technical--No PP Link.
		\item[c] In Experiment~3, this answer was changed to ``because I do not know what data this is about'' in the conditions Non-Technical--PP Link and Non-Technical--No PP Link.
	\end{tablenotes}
	\end{threeparttable}
\end{table*}

\begin{table*}[b]
	\centering
	\small
	\caption*{Expectation of the Website's Data Collection}
	\label{tab:survey-datacollection}
	\begin{threeparttable}
		\begin{tabular}{@{}lrrrrr}
			\toprule
			\multicolumn{6}{l}{Q2: What do you think -- what data does the website [WEBSITE\_NAME] collect about you when you access the website?}\\
			\midrule
			& \textbf{Exp. 1} & \textbf{Exp. 2} & \textbf{Exp. 3} & \textbf{Total} & \textbf{\,\%} \\
		    The posts I am reading on the website.\textsuperscript{R} & 10 & 40 & 17 & 67 & 60.9\,\%\\
	        My residence.\textsuperscript{R} & 6 & 14 & 7 & 27 & 24.5\,\%\\
	        The links I click on the website.\textsuperscript{R} & 14 & 45 & 27 & 86 & 78.2\,\%\\
	        My IP address.\textsuperscript{R} & 11 & 39 & 22 & 72 & 65.5\,\%\\
	        The device I am using to access the website.\textsuperscript{R} & 10 & 36 & 19 & 65 & 59.1\,\%\\
	        The website does not collect any data about its visitors.\textsuperscript{R} & 0 & 4 & 1 & 5 & 4.5\,\%\\
	        My name.\textsuperscript{R} & 2 & 9 & 3 & 14 & 12.7\,\%\\
	        Other websites I visit besides [WEBSITE\_NAME].\textsuperscript{R} & 5 & 17 & 10 & 32 & 29.1\,\%\\
	        Other: [free text] & 3 & 2 & 1 & 6 & 5.5\,\%\\
	        I prefer not to answer. & 0 & 0 & 0 & 0 & 0\,\%\\
			\textbf{\# Answers} & \textbf{61} & \textbf{206} & \textbf{107} & \textbf{374} & \textbf{}\\
			\textbf{\# Participants} & \textbf{16} & \textbf{60} & \textbf{34} & \textbf{110} & \textbf{}\\
			\bottomrule
		\end{tabular}
	\centering
	%\begin{tablenotes}
	%\end{tablenotes}
	\end{threeparttable}
\end{table*}

% \vspace{50mm}

\begin{table*}[h!tb]
	\centering
	\small
	\caption*{Perception of the Cookie Consent Notice Displayed to the Participant}
	\label{tab:survey-notice}
	\begin{threeparttable}
		\begin{tabular}{@{}lrrrrrrrrrrrrrr}
			\toprule
			\multicolumn{15}{l}{This is the cookie consent notice\textsuperscript{a} the website has shown you. [IMAGE]}\\
			\multicolumn{15}{l}{Please rate the following statements about this notice.}\\
			\midrule
			\midrule
			\multicolumn{15}{l}{Q3: I think the number of choices offered by the above cookie consent notice\textsuperscript{b} is ...}\\
			\midrule
			& \textbf{Exp. 1} & \multicolumn{9}{c}{\textbf{Exp. 2}} & \multicolumn{4}{c}{\textbf{Exp. 3}} \\
			& \rot{\textbf{BIN-E1\tnote{c}}} & \rot{\textbf{NOP}} & \rot{\textbf{CON-NN}} & \rot{\textbf{CON-NU}} & \rot{\textbf{BIN-NN}} & \rot{\textbf{BIN-NU}} & \rot{\textbf{CAT-NN}} & \rot{\textbf{CAT-NU}} & \rot{\textbf{VEN-NN}} & \rot{\textbf{VEN-NU}} & \rot{\textbf{TE-PP}} & \rot{\textbf{TE-NP}} & \rot{\textbf{NT-PP}} & \rot{\textbf{NT-NP}} \\
			\midrule
			... too low & 9 & 3 & 3 & 5 & 3 & 1 & 1 & 2 & 1 & 2 & 1 & 1 & 0 & 1 \\
	        ... just right & 7 & 1 & 0 & 3 & 7 & 3 & 2 & 3 & 1 & 2 & 4 & 8 & 6 & 6\\
	        ... too high & 0 & 0 & 1 & 1 & 0 & 0 & 3 & 2 & 0 & 3 & 2 & 0 & 3 & 0 \\
	        ... No answer & 0 & 2 & 0 & 1 & 2 & 0 & 1 & 0 & 1 & 0 & 0 & 0 & 2 & 0\\
	        \textbf{Total} & \textbf{16} & \textbf{6} & \textbf{4} & \textbf{10} & \textbf{12} & \textbf{4} & \textbf{7} & \textbf{7} & \textbf{3} & \textbf{7} & \textbf{7} & \textbf{9} & \textbf{11} &  \textbf{7} \\
	        \midrule
	        \midrule
			\multicolumn{15}{l}{Q4: The above cookie consent notice\textsuperscript{a} allows me to control the website's behavior.}\\
			\midrule
			& \textbf{Exp. 1} & \multicolumn{9}{c}{\textbf{Exp. 2}} & \multicolumn{4}{c}{\textbf{Exp. 3}} \\
			& \rot{\textbf{BIN-E1}} & \rot{\textbf{NOP}} & \rot{\textbf{CON-NN}} & \rot{\textbf{CON-NU}} & \rot{\textbf{BIN-NN}} & \rot{\textbf{BIN-NU}} & \rot{\textbf{CAT-NN}} & \rot{\textbf{CAT-NU}} & \rot{\textbf{VEN-NN}} & \rot{\textbf{VEN-NU}} & \rot{\textbf{TE-PP}} & \rot{\textbf{TE-NP}} & \rot{\textbf{NT-PP}} & \rot{\textbf{NT-NP}} \\
			\midrule
			Strongly disagree & 6 & 3 & 3 & 0 & 2 & 0 & 1 & 1 & 0 & 1 & 1 & 1 & 0 & 0\\
	        Somewhat disagree & 3 & 2 & 0 & 3 & 3 & 2 & 2 & 1 & 1 & 3 & 2 & 0 & 3 & 0\\
	        Neutral & 6 & 0 & 1 & 4 & 3 & 0 & 0 & 1 & 1 & 1 & 1 & 1 & 4 & 2\\
	        Somewhat agree & 1 & 1 & 0 & 2 & 4 & 1 & 4 & 3 & 1 & 1 & 1 & 4 & 4 & 5\\
	        Strongly agree & 0 & 0 & 0 & 0 & 0 & 1 & 0 & 1 & 0 & 1 & 2 & 2 & 0 & 0\\
	        No answer & 0 & 0 & 0 & 1 & 0 & 0 & 0 & 0 & 0 & 0 & 0 & 1 & 0 & 0\\
	        \textbf{Total} & \textbf{16} & \textbf{6} & \textbf{4} & \textbf{10} & \textbf{12} & \textbf{4} & \textbf{7} & \textbf{7} & \textbf{3} & \textbf{7} & \textbf{7} & \textbf{9} & \textbf{11} &  \textbf{7} \\
			\midrule
            \midrule
			\multicolumn{15}{l}{Q5\textsuperscript{b}: I think the decision which option to select in the cookie consent notice\tnote{a} is ...}\\
			\midrule
			& \textbf{} & \multicolumn{9}{c}{\textbf{Exp. 2}} & \multicolumn{4}{c}{\textbf{Exp. 3}} \\
			& \rot{\textbf{}} & \rot{\textbf{}} & \rot{\textbf{}} & \rot{\textbf{}} & \rot{\textbf{}} & \rot{\textbf{}} & \rot{\textbf{CAT-NN}} & \rot{\textbf{CAT-NU}} & \rot{\textbf{VEN-NN}} & \rot{\textbf{VEN-NU}} & \rot{\textbf{TE-PP}} & \rot{\textbf{TE-NP}} & \rot{\textbf{NT-PP}} & \rot{\textbf{NT-NP}} \\
			\midrule
	        ... very easy &&&&&&& 2 & 0 & 1 & 1 & 4 & 3 & 4 & 1\\
	        ... easy &&&&&&& 0 & 2 & 1 & 0 & 0 & 2 & 2 & 2\\
	        ... neither easy nor hard &&&&&&& 2 & 2 & 0 & 2 & 2 & 2 & 5 & 4 \\ 
	        ... hard &&&&&&& 2 & 2 & 1 & 2 & 1 & 1 & 0 & 0\\
	        ... very hard &&&&&&& 1 & 1 & 0 & 2 & 0 & 0 & 0 & 0\\
	         No answer &&&&&&& 0 & 0 & 0 & 0 & 0 & 1 & 0 & 0\\
	        \textbf{Total} & \textbf{} & \textbf{} & \textbf{} & \textbf{} & \textbf{} & \textbf{} & \textbf{7} & \textbf{7} & \textbf{3} & \textbf{7} & \textbf{7} & \textbf{9} & \textbf{11} &  \textbf{7} \\
			\midrule
			\bottomrule
		\end{tabular}
	\centering
	\begin{tablenotes}
			\item[a] In Experiment~3, ``cookie consent notice'' was changed to ``privacy notice'' in the conditions Non-Technical--PP Link and Non-Technical--No PP Link.
			\item[b] Q5 was only shown to participants who had seen a category- oder vendor-based notice on the website.
			\item[c] BIN-E1 = the binary notice shown at six different positions in Experiment~1; NOP = no option; CON = confirmation; BIN = binary; CAT = categories; VEN = vendors; NN = non-nudging; NU = nudging; TE = technical; NT = non-technical; PP = privacy policy link; NP = no privacy policy link.
	\end{tablenotes}
	\end{threeparttable}
\end{table*}

\begin{table*}[h!tb]
	\centering
	\small
	\caption*{Perception of the Cookie Consent Notice Displayed to the Participant (cont.)}
	\label{tab:survey-notice2}
	\begin{threeparttable}
		\begin{tabular}{@{}llrrrr}
			\toprule

			\multicolumn{6}{l}{Q6\tnote{a}: Please explain your answer to the previous question. [free text answers, coded by two authors]}\\
			\midrule
			\textbf{Code} & \textbf{Explanation} & \textbf{Exp. 2} & \textbf{Exp. 3} & \textbf{Total} & \textbf{\%} \\
			Transparent & The participant considers the consent notice to be transparent. & 1 & 5 & 6 & 15.8\,\%\\ 
			Privacy & The participant's preferences are privacy-focused, \ie, the least invasive option is chosen. & 2 & 5 & 7 & 18.4\,\%\\
			Options clear & The options offered by the consent notice are considered clear / easy to understand. & 0 & 3 & 3 & 7.9\,\%\\
			Options unclear & The options offered by the consent notice are considered unclear / not easy to understand. & 4 & 2 & 6 & 15.8\,\% \\
			Notice clear & The participant expressed that the mechanism was clear but did not specify which part. & 1 & 3 & 4 & 10.5\,\%\\
			Notice unclear & The participant expressed that the mechanism was unclear but did not specify which part. & 2 & 0 & 2 & 5.3\,\%\\
			Too complicated & The consent notice was considered too complex. & 4 & 1 & 5 & 13.2\,\%\\
			Don't care & The participant stated they did not care which cookies the website used. & 3 & 0 & 3 & 7.9\,\%\\
			Other & & 4 & 2 & 6 & 15.8\,\%\\
			\midrule

			\textbf{\# Participants} &  & \textbf{60} & \textbf{34} & \textbf{94} & \textbf{}\\
			\bottomrule
		\end{tabular}
	\centering
	\begin{tablenotes}
			\item[a] Q6 was only shown to participants who had seen a category- oder vendor-based notice on the website.
	\end{tablenotes}
	\end{threeparttable}
\end{table*}

\begin{table*}[h!tb]
	\centering
	\small
	\caption*{General Understanding of Cookie Consent Notices}
	\label{tab:survey-understanding}
	\begin{threeparttable}
		\begin{tabular}{@{}llrrrrr}
			\toprule
            \multicolumn{7}{l}{This is another cookie consent notice. [Image of the binary notice in Figure~\ref{fig:cookie_banners} (a) (bb)]}\\
            \midrule
            \midrule
			\multicolumn{7}{l}{Q7: What do you think happens when you click ``Decline''? [free text answers, coded by two authors]}\\
			\midrule
			\textbf{Code} & \textbf{Explanation} & \textbf{Exp. 1} & \textbf{Exp. 2} & \textbf{Exp. 3} & \textbf{Total} & \textbf{\%} \\
			\midrule
			Site blocked & The content of the website cannot be accessed at all. & 6 & 13 & 9 & 28 & 29.8\,\%\\
			Functionality limited & The content of the website can be viewed, but some parts may not work. & 2 & 10 & 5 & 17 & 18.1\,\%\\
			Site accessible & The content of the website can be accessed. & 0 & 3 & 1 & 4 & 4.3\,\%\\
			No data collected & The website visitor's personal data is not collected or processed. & 2 & 4 & 5 & 11 & 11.7\,\%\\
			No cookies set & The website does not store any cookies in the visitor's browser. & 1 & 8 & 3 & 12 & 12.8\,\%\\
			Less ads & The website displays less or no ads. & 0 & 3 & 2 & 5 & 5.3\,\%\\
			Notice & The participants only mentions effects regarding the consent notice. & 0 & 2 & 3 & 5 & 5.3\,\%\\
			No change & Declining cookies does not have any effect. & 4 & 7 & 1 & 12 & 12.8\,\%\\
			Don't know & & 2 & 0 & 1 & 3 & 3.2\,\%\\
			Other & & 0 & 2 & 4 & 6 & 6.4\,\%\\
			\midrule
			\textbf{\# Participants} &  & \textbf{15} & \textbf{51} & \textbf{28} & \textbf{94} & \textbf{}\\
			\midrule
			\midrule
			\multicolumn{7}{l}{Q8: What do you think happens when you click ``Accept''? [free text answers, coded by two authors]}\\
			\midrule
			\textbf{Code} & \textbf{Explanation} & \textbf{Exp. 1} & \textbf{Exp. 2} & \textbf{Exp. 3} & \textbf{Total} & \textbf{\%} \\
			\midrule
			Data collected & The participant's personal data is collected and/or processed. & 9 & 10 & 10 & 29 & 30.9\,\%\\
			Cookies stored & Cookies are stored in the user's browser. & 4 & 9 & 6 & 19 & 20.1\,\%\\
			Site accessible & The content of the website can be accessed. & 0 & 16 & 5 & 21 & 22.3\,\%\\
			Notice & The participants only mentions effects regarding the consent notice. & 0 & 3 & 2 & 5 & 5.3\,\%\\
			Ads & The participant is subject to advertising. & 6 & 11 & 6 & 23 & 24.5\,\%\\
			Profiling & The participant's personal data is used to create a profile of their interests. & 5 & 8 & 6 & 19 & 20.2\,\%\\
			Other purposes & The participant's personal data is used for other purposes. & 2 & 0 & 2 & 4 & 4.3\,\%\\
			No change & Clicking ``Accept'' does not have any effect. & 0 & 4 & 3 & 7 & 7.4\,\%\\
			Don't know & & 0 & 3 & 0 & 3 & 3.2\,\%\\
			Other & & 0 & 3 & 1 & 4 & 4.3\,\%\\
			\midrule
			\textbf{\# Participants} &  & \textbf{15} & \textbf{51} & \textbf{28} & \textbf{94} & \textbf{}\\
			\bottomrule
		\end{tabular}
	\centering
	%\begin{tablenotes}
	%\end{tablenotes}
	\end{threeparttable}
\end{table*}

\end{document}